\documentclass[a4paper,fleqn,usenatbib]{mnras}

\usepackage{newtxtext,newtxmath}

\usepackage[T1]{fontenc}
\usepackage{ae,aecompl}
\usepackage{afterpage}

\usepackage{graphicx}	
\usepackage{amsmath}	
\usepackage{amssymb}	
\usepackage{textcomp}
\usepackage{subfig}

\title[A time domain simulator for JWST]{JexoSim: A time domain simulator of exoplanet transit spectroscopy with JWST}

\author[S. Sarkar et al.]{
Subhajit Sarkar,$^{1}$\thanks{E-mail: subhajit.sarkar@astro.cf.ac.uk (S.S.)}
Nikku Madhusudhan$^{2}$ and Andreas Papageorgiou$^{1}$
\\
$^{1}$School of Physics and Astronomy, Cardiff University, The Parade, Cardiff, CF24 3AA, UK\\
$^{2}$Institute of Astronomy, University of Cambridge, Madingley Road, Cambridge, CB3 0HA, UK\\
}

\date{Accepted XXX. Received YYY; in original form ZZZ}

\pubyear{2019}

\begin{document}
\label{firstpage}
\pagerange{\pageref{firstpage}--\pageref{lastpage}}
\maketitle

\begin{abstract}
The James Webb Space Telescope (JWST) will perform exoplanet transit spectroscopy in the coming decade promising transformative science results.  All four instruments on board can be used for this technique which reconstructs the atmospheric transmission or emission spectrum of an exoplanet from wavelength-dependent light curve measurements.  Astrophysical and instrumental noise and systematics can affect the precision and accuracy of the final  spectrum, and hence, the atmospheric properties derived from the spectrum. Correlated noise and time-dependent systematics that can bias the measured signal must be accounted for in the final uncertainties. However quantifying these effects can be difficult with real data or simple analytic tools. Existing publicly-available simulators for JWST do not adequately simulate complex time domain processes on exoplanetary transit observations. We report JexoSim, a dedicated time domain simulator for JWST including all four instruments for exoplanet transit spectroscopy. JexoSim models both the astrophysics and the instrument, generating 2-D images in simulated time akin to a real observation. JexoSim can capture correlated noise and systematic biases on the light curve giving it great versatility. Potential  applications of JexoSim include performance testing of JWST instruments, assessing science return, and testing data reduction pipelines. We describe JexoSim, validate it against other simulators, and present examples of its utility. 
\end{abstract}

\begin{keywords}
infrared: planetary systems  -- stars: activity -- space vehicles: instruments
\end{keywords}


\section{Introduction}
\label{Intro}
Exoplanet science has progressed rapidly over the past two decades with over 4000 exoplanets now known. The technique of transit and eclipse spectroscopy \citep{Seager2000,Charb2002,Tinetti2007,Deming2013} has been used to further characterise selected exoplanets permitting constraints on their atmospheric properties such as chemical compositions, temperature-pressure profiles, and presence of clouds/hazes \citep[e.g.][]{Sing2016,Kreidberg2014,Madhusudhan2019}. A transmission spectrum, observed during primary transit, probes the day-night terminator region of the atmosphere. On the other hand, an emission spectrum, observed at secondary eclipse, probes the dayside atmosphere of the planet. 

A transit or eclipse spectrum is reconstructed from the wavelength-dependent light curves observed during the primary transit or secondary eclipse event. The light curve in each spectral bin provides the transit or eclipse depth at that wavelength. The spectrum, $p(\lambda)$, is represented by the fractional transit or eclipse depth vs wavelength with the uncertainty on $p(\lambda)$, $\sigma_p(\lambda)$, represented by the error bars. In primary transit this gives the transmission spectrum of the planet atmosphere at the day-night terminator: $p(\lambda) = {R_p(\lambda)}^2/R_s^2$, where $R_p(\lambda)$ is the apparent planet radius, and $R_s$ is the stellar radius.  In secondary eclipse it gives the planet dayside emission spectrum: $p(\lambda) = F_p(\lambda)/(F_p(\lambda) + F_s(\lambda)) \approx F_p(\lambda)/F_s(\lambda)$, where $F_p(\lambda)$ is the flux from the planet and $F_s(\lambda)$ is the flux from the star. $p(\lambda)$ is usually obtained through the fitting of model light curves (e.g. the formulation by \cite{Mandel2002} which includes the effects of stellar limb darkening) to the colour-dependent light curves.  The final spectrum will have a given accuracy (i.e. a measure of statistical bias or how close the mean value of a large set of measurements of $p(\lambda)$ is from the true value), and precision (i.e. a measure of the statistical random variability, e.g., the standard deviation of the distribution of the measured values of $p(\lambda)$ around its mean). 

The error bars on a spectrum should reflect the overall final uncertainty on the measured value of $p(\lambda)$, both in terms of the precision and accuracy. Accuracy is vulnerable to processes that distort the measured light curve such that the fitted value for $p(\lambda)$ becomes biased from the true value. Examples are stellar surface inhomogenities such as star spots and faculae \citep{Rackham2018} or instrumental systematics such as detector persistence \citep{Berta2012}. Although such systematics can be detrended in data processing, the corrections may not be perfect and residual biases may persist. The precision in turn is vulnerable to multiple sources of statistical noise, both astrophysical and instrumental. Such noise can be uncorrelated (`white noise') or correlated (`red noise'), and it can be Gaussian or non-Gaussian.  Detecting and accounting for the uncertainty arising from correlated noise can be difficult, leading to possible underestimation of the error bar \citep{Pont2006}. Sources of correlated noise include astrophysical sources such as stellar pulsation and granulations \citep{Sarkar2018} or instrumental sources such as pointing jitter.  The size of the error bar impacts directly on the spectral retrieval process \citep[e.g.][]{Madhusudhan2009, Line2013, Waldmann2015, Madhusudhan2018}, by which forward models are fitted to the data in a Bayesian framework, and the best fitting model obtained with its own associated uncertainty. Therefore correctly accounting for the final error on the spectrum, including correlated noise and any residual biases, is crucial to the atmospheric properties retrieved from the spectrum.

The James Webb Space Telescope (JWST) \citep{Gardner2006} is an optical-infrared space observatory in the final stages of ground development, projected for launch in 2021, which promises transformative science in the characterisation of exoplanet atmospheres.  The primary mirror has a total collecting area of 25 m$^2$, the largest of any space telescope to date.  All four of its instruments will be capable of transit spectroscopy of exoplanets.  These are: Near Infrared Imager and Slitless Spectrograph (NIRISS) \citep{Doyon2012}, Near Infrared Camera (NIRCam) \citep{Beichman2012}, Near Infrared Spectrograph (NIRSpec) \citep{Ferruit2014}, and Mid-Infrared Instrument (MIRI)  \citep{Rieke2015}. JWST is expected to conduct detailed atmospheric spectroscopy of known exoplanets to answer a range of science questions \citep{Bean2018}. Therefore, in the lead up to launch and operations, the exoplanet community will benefit from simulation tools that will permit the performance of JWST and its instruments to be assessed and that will facilitate formulation of optimal observing strategies. 

Currently, there is a need for a time domain simulator for transit spectroscopy with JWST. Recently, the Pandexo simulation package \citep{Batalha2017} has been developed for simulating JWST observations of transit spectroscopy. Pandexo is a radiometric simulator, that obtains signal and noise estimates using parametric models for all four JWST instruments. Radiometric simulators have the advantage of computational efficiency to provide rapid simulations of observations. However, since such simulators do not model the time domain directly in a frame-by-frame simulation, they are limited in the ability to estimate the effects of complex time-dependent systematics and time-correlated random noise. On the other hand, by modelling the time domain using a dynamic numerical simulation, evolving with small time steps, it is possible to include such processes, and measure their impact on accuracy and precision directly. By simulating the transit light curve within such a simulation, a simulated observation with time series spectral images can be produced. These can be used to measure experimental uncertainties and biases from a wide range of noise sources and systematics directly on the final reconstructed planet spectrum.  Such simulated data can also be used in the validation of data reduction pipelines, as well as the verification of calibration strategies, optimization of observing modes, and testing of data management strategies.  

In this work, we report JexoSim, a dedicated time domain simulator for JWST.  JexoSim simulates the planet transit or eclipse generating 2-D images in simulated time,
complete with multiple sources of noise and systematics, and then passes these to an associated pipeline to process the results. JexoSim has its origins in the generic time domain code, ExoSim \citep{Sarkar2016}, but is an independent, dedicated development that has been modified and optimised for JWST. This has included increased modularity and adaptations for the many JWST instrument modes.  ExoSim itself has been extensively used during the Phase A study of the European Space Agency Cosmic Vision M4 mission ARIEL (Atmospheric Remote-sensing Exoplanet Large-survey) \citep{Tinetti2018}.  
JexoSim is currently not open-source but we plan to make it accessible to the exoplanet community in the future.

In this paper we first describe JexoSim, its algorithm, associated pipeline and types of results.
 We then validate JexoSim by comparing its results to those published for Pandexo. We first compare its photon conversion efficiency profiles with those from Pandexo. We then compare photon noise results from JexoSim with those presented in \cite{Batalha2017} for Pandexo and independent simulators for individual instruments: the NIRISS SOSS 1-D simulator developed by the NIRISS instrument team\footnote{http://maestria.astro.umontreal.ca/niriss/simu1D/simu1D.php}, the NIRSpec Exoplanet Exposure Time Calculator (NEETC) \citep{Nielsen2016}, and a 1-D simulator developed for NIRCam and MIRI \citep{Greene2016}.  Next, we demonstrate JexoSim's capability to produce a `noise budget' for JWST instruments observing the super-Earth GJ 1214 b. Finally, we use JexoSim to simulate a transit spectroscopy observation of the hot Jupiter HD 209458 b, producing a transmission spectrum with error bars.  We aim to show the potential versatility and unique capabilities of JexoSim applied to JWST.

\section{Previous simulators} 
 
Several studies have previously evaluated JWST transit spectroscopy performance using simulators. \cite{Batalha2015} reported a JWST NIRSpec simulator, which included Poisson noise, jitter drift, flat field errors, read noise and zodiacal light. It also performed convolutions with the point-spread function and the intra-pixel response function.  The capability of this simulator for producing time series images to investigate complex time-dependent processes has not been explained in detail. While not publicly available, results from it were used in \cite{Beichman2014}, which comprehensively evaluated the science opportunities and potential of all four JWST instruments. \cite{Beichman2014} also includes results from the NIRISS SOSS 1-D Simulator. This simulator includes Poisson noise, read out noise, an optional noise floor, and assumes extraction is performed optimally. Since a 1-D simulator may not capture all effects on a 2-D image (especially as the spectral trace is curved in NIRISS SOSS) a 2-D simulator is also being developed by the same team.   
The NIRSpec instrument performance simulator \citep{Piqueras2008} simulates all four NIRSpec modes in 2-D simulations that capture a number of effects on the focal plane image.  The NIRSpec Exoplanet Exposure Time Calculator (NEETC) \citep{Nielsen2016} is an instrument simulator for NIRSpec based on a radiometric model, modeling nine instrument configurations. The noise is calculated through equations that compute the read out noise, kTC noise and Poisson noise from the stellar target and dark current.  Simulators for NIRCam include PyNRC \citep{Leisenring2018}, which produces 2-D images. \cite{Greene2016} performed simulations of NIRISS SOSS, NIRCam and MIRI low-resolution spectroscopy mode (LRS) using a 1-D simulation tool that calculates the signal and noise per spectral element using equations. The noise sources included photon noise from the target system, instrumental emission and dark current, as well as read noise and a noise floor. All the tools discussed thus far do not utilize dynamical time domain simulation to obtain their results.

Most recently, as discussed in Section \ref{Intro}, Pandexo has been used for JWST simulation.  Pandexo utilises the Pandeia engine \citep{Pontoppidan2016} which is also used in the  the JWST Exposure Time Calculator. As described in \cite{Batalha2017}, Pandeia
incorporates up-to-date instrument parameters including throughputs, PSFs, saturation levels, correlated read noise and flat field errors for all JWST instruments. Advantages of Pandexo over other simulators include that it simulates all JWST instruments and modes, and has a very accessible web interface. It can also generate 2-D spectral images. As described in \cite{Batalha2017}, Pandexo models pure shot noise, read noise and background noise (zodiacal light and cirrus) to obtain signal and noise estimates for a transit observation without creating the full light curve.  Other detector systematic noise and jitter noise are not included, but a noise floor can be included.  It utilizes last-minus first (LMF) processing as its standard procedure for obtaining the final integration signal.  The overall formulation gives rapid results making it useful as a community tool for quickly obtaining a first order performance estimate for a proposed JWST observation. However limitations exist since it does not simulate all potential noise sources and systematics, and there is no frame-by-frame simulation of the time domain.

In the context of these previous simulators, JexoSim attempts to provide a unique capability. It can simulate all four JWST instruments in transit or eclipse spectroscopy.  It generates signal and noise using a dynamical approach, rather than a static one.
It can model the time domain, creating full light curves. It can generate simulated observations consisting of 2-D spectral time series images, which can be used as input to data reduction pipelines. It can capture the effects of time-dependent correlated noise and systematics directly on the exoplanet spectrum.  As we learn more deeply about complex noise sources and systematics affecting the various JWST instruments, JexoSim has the potential to include these effects so that its simulations will adapt to match future real observations.  With these capabilities, JexoSim has the potential to be a highly versatile JWST simulator.

\section{JexoSim}

In this section we describe the JexoSim framework and algorithm. JexoSim utilizes a modular structure which approximates the flow of information from the star to the detector, outputting time series spectral images, which then flow into a data reduction pipeline to produce the final results for the user.  JexoSim is written mainly in Python 2.7, with two configuration files written in XML which are the primary inputs.  The `common' configuration file controls the telescope parameters and also controls the exosystem, observation time, which noise sources and backgrounds to simulate, as well as which instrument modes to call on.  There are also options for which kind of results to return from the pipeline.  The second `instrument' configuration file is for the specific instrument mode selected, and can be interchanged depending on which instrument and mode is desired.  Different instrument modes thus have their own dedicated configuration files, permitting flexibility in adding future modes. 

By default, JexoSim directly feeds its time series image data into a processing pipeline, the JexoSim Default Pipeline (JDP), which then delivers the desired results as outputs.  However the JDP can be replaced by other pipelines to test and validate data reduction and processing steps, and JexoSim can deliver its output as FITS files to facilitate this.
The ability to model 2 dimensional spectral traces was also specifically developed for JexoSim as this is important for NIRISS which has a curved spectral trace.  The code is highly modularised to allow for future modification and upgrading.
  As well as standard Python packages, the current version of the code also utilizes several open source Python packages which are listed in the Acknowledgements. In addition, the code uses the Open Exoplanet Catalogue (OEC) \citep{Rein2012}\footnote{http://www.openexoplanetcatalogue.com/} and PHOENIX  BT-Settl spectra \citep{Allard2012}\footnote{https://phoenix.ens-lyon.fr/Grids/BT-Settl/CIFIST2011\_ 2015/FITS/} databases. The exosystem can be either be user-defined with relevant parameters adjusted in the `common' configuration file, or if a known exoplanet is inputed by name, all required parameters are obtained via the OEC database. For this paper, we modeled the following instrument modes: NIRISS SOSS (order 1), NIRCam (long-wave channel) F444W, NIRSpec G395M/290LP and MIRI LRS slitless mode.

\begin{figure}
 \begin{center}
 	\includegraphics[trim={0cm 1cm 0cm 2.5cm}, clip, width=0.97\columnwidth]{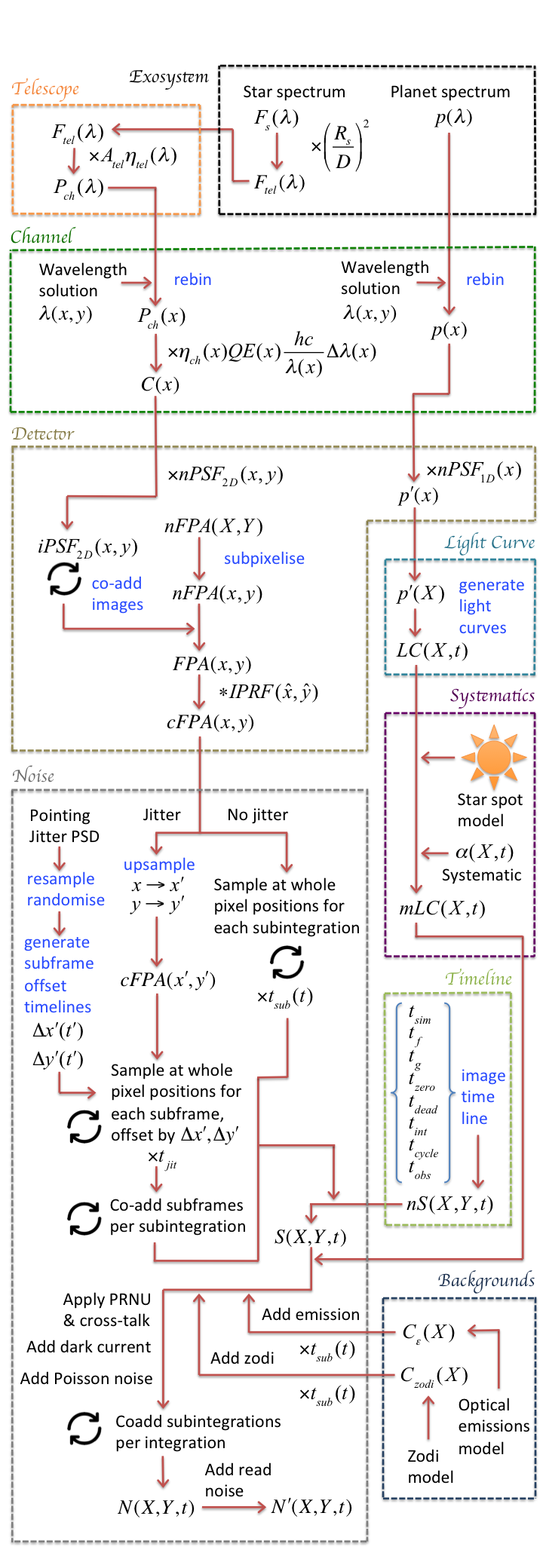}
 \end{center}
\vspace{-0.4em}  
    \caption{Major elements in the JexoSim algorithm.  Abbreviations and steps are described in Section \ref{algo_section}. Boxes represent functional modules in the code.}
    \label{fig:Architecture}
\end{figure}

\subsection{JexoSim algorithm}
\label{algo_section}

In this section we summarize the main steps in the JexoSim algorithm (Figure \ref{fig:Architecture}). 
 
\subsubsection{Star to telescope}

The algorithm begins with a high resolution model stellar flux density spectrum, $F_{s}(\lambda)$ (in units of W/m$^2$/\textmu m).  The  $Exosystem$ module instantiates the star and planet as object classes, and the best matching PHOENIX  BT-Settl model is selected depending on the host star parameters for temperature, metallicity and log $g$.  The flux density at the telescope, $F_{tel}(\lambda)$, for an exosystem at distance $D$ is then given by:
\begin{equation}
F_{tel}(\lambda)=F_{s}(\lambda)(R_s/D)^2
\end{equation}
 
\vspace{0.1cm} 
 
\subsubsection{OTE}

$F_{tel}(\lambda)$ is fed into the $Telescope$ module, where it is multiplied by the effective collecting area of the telescope aperture, $A_{tel}$, to give the radiant power per wavelength (in units of W/\textmu m) entering JWST. The light then passes through the Optical Telescope Element (OTE), where is it modulated by the transmission of each mirror surface, giving a combined transmission, $\eta_{tel}(\lambda)$.  For this paper we utilize the OTE throughput file from the Pandeia v1.3 database\footnote{http://ssb.stsci.edu/pandeia/engine/1.3/pandeia\_data-1.3.tar.gz} \citep{Pontoppidan2016}  
which contains the combined estimated end-of-life throughput of the four mirrors in the OTE (primary, secondary, tertiary and fine steering mirror). The radiant power per wavelength, $P_{ch}(\lambda)$ entering the instrument channel is thus given by:
\begin{equation}
P_{ch}(\lambda) = F_{tel}(\lambda)A_{tel}\eta_{tel}(\lambda)
\end{equation}

\vspace{0.1cm}

\subsubsection{Instrument channel}

The light then passes into the selected instrument and channel, configured for a particular observing mode, and is dispersed into a spectrum by either a grism or prism.  This is simulated in the $Channel$ module.  This dispersion is modeled using a wavelength solution $\lambda(X,Y)$, where $\lambda$ is a function of the discrete variables $X$ and $Y$ which are coordinates representing pixels in the 2-D focal plane detector array $FPA(X,Y)$ consisting of $N_X \times N_Y$ pixels ($X= 0,1,2,...,N_X-1$, $Y = 0,1,2,...,N_Y-1$).
In most cases we assume the spectral trace is parallel to the $X$-axis (spectral or dispersion axis); however this is not so for NIRISS where there is a curved trace.  Wavelength solutions of this type  were calculated for each channel from data in the corresponding dispersion files in the Pandeia v1.3 database.  In addition for NIRISS, the database gave trace files from which the $Y$ position of the spectral trace as a function of $X$ could be obtained. To ensure we Nyquist sample the stellar signal within the simulation, each pixel is subdivided  by a factor of 3 in each axis into subpixels. Therefore we utilise a 2-D subpixelised focal plane array which initially has `null' (zero) values, $nFPA(x,y)$, where $x$ and $y$ are discrete subpixel coordinates
($x = 0,1,2,...,3N_X-1$, $y = 0,1,2,...,3N_Y-1$).
The wavelength solution is resampled to to give $\lambda(x,y)$.  $P_{ch}(\lambda)$ is rebinned to the wavelength solution for each subpixel column to give $P_{ch}(x)$.   The transmissions of each optical element (e.g. mirrors, filters, grisms) in the instrument channel modulate the stellar signal having a combined transmission, $\eta_{ch}(\lambda)$, rebinned to the wavelength solution to give $\eta_{ch}(x)$. Again we utilize the the throughput files given in the Pandeia v1.3 database which generally have separate files for the dispersing element, filters and internal optics.
For NIRISS SOSS we apply in addition a 33\% reduction in the transmission due to undersizing of the grism with respect to the pupil\footnote{http://jwst.astro.umontreal.ca/?page\_id=51}. For MIRI LRS, we assumed seven internal mirrors each with a transmission of 0.982, and a contamination factor of 0.8 as used in the Pandeia v1.3 configuration file for MIRI LRS.

\subsubsection{Convolution with PSF}

Convolution with the point spread function (PSF) to generate the spectral image is simulated as follows. A 2-D PSF array, $nPSF_{2D}(x,y)$, is generated for each subpixel column wavelength, $\lambda(x)$. This is a `null' array in that it is normalized to a total volume of unity.  These PSFs were generated for each channel using WebbPSF v0.8.0\footnote{https://github.com/spacetelescope/webbpsf} \citep{Perrin2012}. 
 In WebbPSF, each PSF was oversampled by a factor of 3 to make it compatible with the default oversampling rate for the JexoSim focal plane image, normalised to a total volume of unity and produced without any Gaussian blur simulating jitter.  JexoSim can independently generate PSFs from Airy functions or Gaussian functions; however PSFs from WebbPSF will be more realistic as they will incorporate known aberrations from wavefront errors.
In the case of NIRISS SOSS, the PSFs generated by WebbPSF are elongated in the cross-dispersion direction, which is due to the GR700XD grism using a cylindrical lens \citep{Goudfrooij2015}.
Next, the modified stellar spectrum is multiplied by $\Delta \lambda (x)$, the wavelength span across each subpixel column, giving the radiant power falling on each subpixel column (in units of W). The wavelength-dependent quantum efficiency (QE) of the detector, is applied at this stage  as a wavelength dependent array, $QE(x)$ (based on Pandeia v1.3 database QE files for each instrument).  The conversion to photoelectrons is also performed, to give a spectrum of count rates per subpixel column, $C(x)$, in units of e$^-$/s:
\begin{equation}
C(x) = P_{ch}(x)\eta_{ch}(x)QE(x)\frac{hc}{\lambda(x)}\Delta \lambda(x)
\end{equation}
\noindent where $h$ and $c$ are Planck's constant and the speed of light respectively. 
The algorithm then continues in the $Detector$ module.  Here, each `null' PSF, $nPSF_{2D}(x,y)$, is multiplied by the corresponding value of $C(x)$ to give the 2-D image, $iPSF_{2D}(x,y)$, centred on the subpixel column, $x$, and subpixel row, $y$.  Therefore:
\begin{equation}
iPSF_{2D}(x,y) = C(x)nPSF_{2D}(x,y)
\end{equation}
\noindent For all channels other than NIRISS, $y$ is a constant value, but for NIRISS it is a function of $x$.  Since for non-coherent imaging systems, image formation is linear in power, the final spectral image can be generated by simple coaddition of the individual images for each subpixel column.  This coaddition is used to generate the focal plane detector array, $FPA(x,y)$, containing the stellar spectral image in units of e$^-$/s:
\begin{equation}
 FPA(x,y) = \sum_{x} {iPSF_{2D}(x,y)}
\end{equation} 

\subsubsection{Intra-pixel response function} 

\begin{figure}
\begin{center}
    \includegraphics[trim={0 0 0 0}, clip,width=1.0\columnwidth]{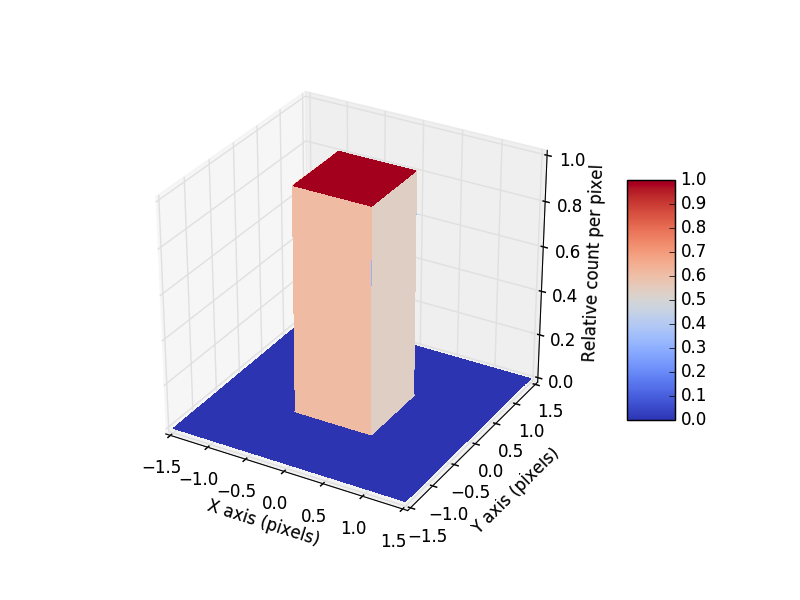}
{A: $l_d$ $\sim$ 0 \textmu m}
\end{center}

\begin{center}
 \includegraphics[trim={0 0 0 0}, clip,width=1.0\columnwidth]{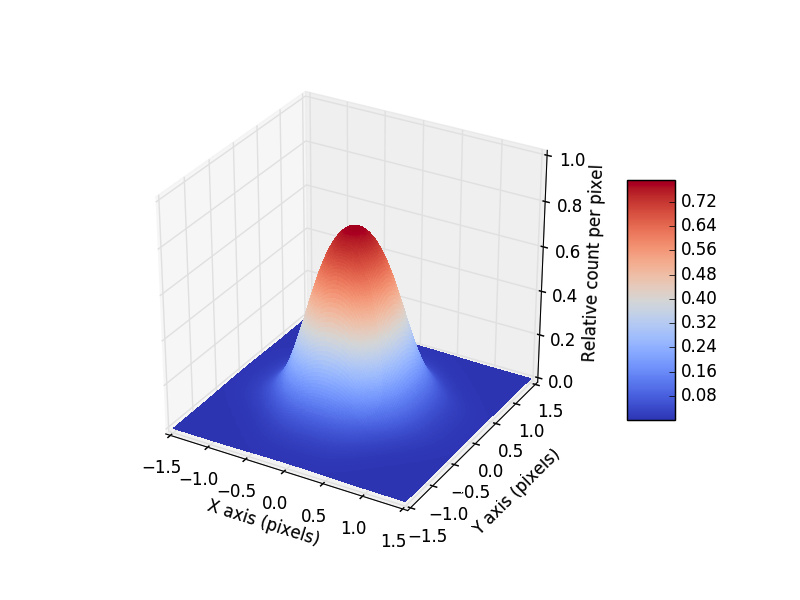}
{B: $l_d$ = 3.7 \textmu m}
\end{center}
    
\caption{Intra-pixel response functions generated by JexoSim. A: `Top hat' function. B: Bell-shaped function with significant pixel crosstalk.}
\label{IPRF}  
\end{figure}

\begin{figure}
 \begin{center}
 	\includegraphics[trim={0cm 0.0cm 0cm 0cm}, clip, width=1.0\columnwidth]{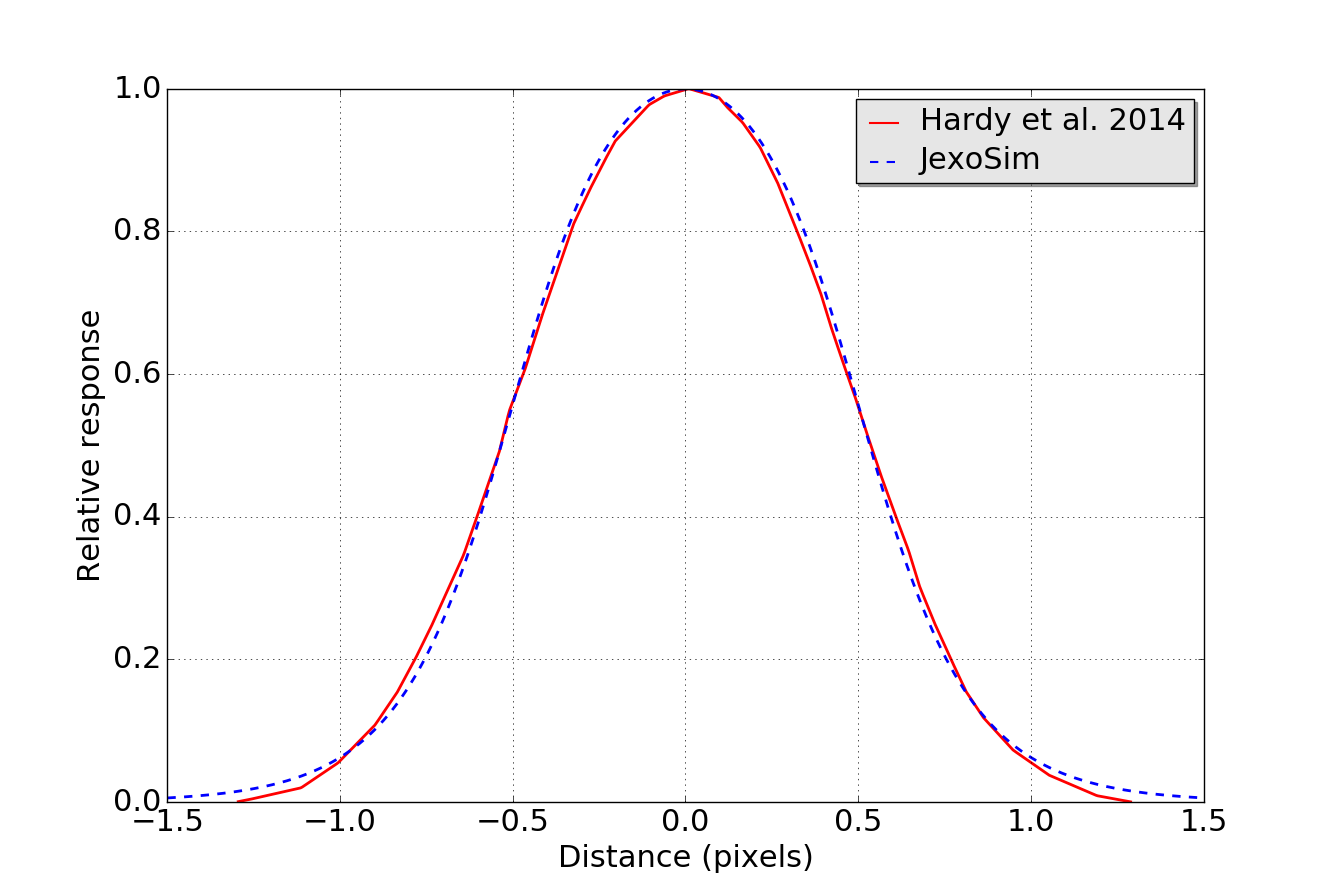}
 \end{center}

    \caption{Cross-sections through intra-pixel response functions compared.  JexoSim IPRF is generated with $l_d$ of 3.7 \textmu m.  This approximates the bell-shaped response function measured in \protect\cite{Hardy2014}.  The IPRF from \protect\cite{Hardy2014} is given for a pixel at 940 nm. It was truncated to exclude negative values. Both IPRFs were normalized to a maximum of unity.}
    \label{fig:IPRF3}
     
\end{figure}

Next we take into account variations in intra-pixel response and inter-pixel cross-talk by the generation of an intra-pixel response function (IPRF). JexoSim performs a 2-D convolution between the IPRF (normalized to produce a volume of unity) and $FPA(x,y)$ to generate a grid $cFPA(x,y)$ such that each point sampled on $cFPA(x,y)$ gives the count over a region covered by the IPRF. In the absence of pixel-to-pixel optical cross-talk, this would be the count over a single pixel. Different positions on the grid capture the variation in the pixel count due to shifts in the image, and this  facilitates the pointing jitter simulation described below.
\cite{Hardy2014} examined the intra-pixel response of H1RG (5 \textmu m cutoff) detectors for JWST.  The pixel response variation was mostly attributed to diffusion of the signal to neighbouring pixels, with a bell-shaped response function with broader wings than a Gaussian.  JexoSim currently utilizes an intra-pixel response function developed for ExoSim which is a 2-D generalization of the 1-D model from \cite{Pascale2015}, itself based on \cite{Barron2007}. If we let $\kappa$ = $\Delta_{pix}/2$, where $\Delta_{pix}$ is the pixel pitch:
\begin{equation}
\begin{split}
IPRF(\hat{x},\hat{y}) = 
\left[\tan^{-1}{\left(\tanh \frac{\kappa-\hat{x}}{2l_d}\right)}-\tan^{-1}{\left(\tanh \frac{-\kappa-\hat{x}}{2l_d}\right)}\right] \cdot\\
\left[\tan^{-1}{\left(\tanh \frac{\kappa-\hat{y}}{2l_d}\right)}-\tan^{-1}{\left(\tanh \frac{-\kappa-\hat{y}}{2l_d}\right)}\right] {\ }\\
\end{split}
\end{equation}
where $l_d$ is the diffusion length, which characterizes the fall off of response at edges of the pixel, and
$\hat{x}$ and $\hat{y}$ are distances from the pixel centre in the same units as $\Delta_{pix}$ and $l_d$.  To ensure the intra-pixel response function is Nyquist sampled prior to convolution with the focal plane array, the 2-D array representing the IPRF is generated at a higher spatial frequency than the focal plane image. The convolution is performed in Fourier space by multiplying the Fourier transform of the IPRF (normalized to a volume of unity) with the Fourier transform of the image interpolated to the same spatial frequency.  The product is then downsampled in Fourier space back to the original spatial frequency of the image before the inverse Fourier transform to produce the convolved focal plane, $cFPA(x,y)$. If $l_d$ is set to $\sim$ 0 it will approach a `top hat' function within the confines of the whole pixel area, and thus with no inter-pixel diffusion (Figure \ref{IPRF} A).  If $l_d$ is set to 3.7 \textmu m we generate a `bell-shaped' function (Figure \ref{IPRF} B), which reasonably approximates the profile from \cite{Hardy2014} (Figure \ref{fig:IPRF3})\footnote{We used Webplotdigitizer \citep{Webplot} to extract all comparison data from reference sources used for this paper.}.  For this study however we have elected to use the top hat function since the bell-shaped function introduces significant charge diffusion contributing to pixel cross-talk.  This would in turn require a cross-talk correction step in data reduction which we have not yet developed.  Simulating the effect of cross-talk is complicated by differences in the baseline QE between pixels, i.e. the photo-response non-uniformity (PRNU), since the electrons diffusing into a pixel (and contributing to its final count) will arise from neighbouring pixels, each with slightly different QEs.
Thus different neighbouring pixels will contribute different amounts of crossed electrons depending on their responsivity to light.  The way this effect can be simulated is described in Section \ref{sec:PRNU}. Future versions of JexoSim could utilize a bell-shaped IPRF together with its corresponding pipeline step.

\vspace{6pt}

\subsubsection{Timing} 

In JWST terminology \citep{Rauscher2007}, a frame is the unit of data that results from sequentially clocking through and reading out a rectangular area of pixels (i.e a subarray). The time taken to read out the subarray is the frame time, $t_f$. 
Between resets of the detector, signal accumulates in the subarray over time as an integration ramp. In the MULTIACCUM read out pattern \citep{Rauscher2007}, the frames that fall within the time period of this ramp are divided into $n$ equally-spaced groups, each consisting of $m$ frames.  A group is a thus a set of one or more consecutively read frames with no intervening resets, and for exoplanet time series modes $m=1$ \citep{Batalha2017}.  The groups are sampled non-destructively, i.e. return non-destructive reads (NDRs) up the ramp.  The group time, $t_g$, is the time interval between reads of the same pixel in the first frame of successive groups \citep{Rauscher2007}.  In JWST terminology, an integration starts with a reset followed by a series of non-destructive reads up the ramp, and an exposure is the complete series of consecutive integrations making up an observation \citep{Nielsen2016, Batalha2017}.

In JexoSim, we assume $m=1$ and $n$ NDRs (corresponding to groups) up the ramp per integration cycle.  The `integration cycle' is defined to be the interval between detector resets, and includes the integration ramp and `dead time' taken up in idling and resetting of the detector. The time interval between succesive NDRs is given by $t_g$, which is user-defined. The time duration of the initial (`zeroth') NDR, $t_{zero}$, is separately user-defined, as is the dead time, $t_{dead}$. The dead time by default is assumed  to precede the integration ramp, but if needed it can divided into pre- and post- ramp components. If all the cycle elements are to be equally-spaced, then $t_{zero}$ and $t_{dead}$ can be set to equal $t_g$.
JexoSim uses the frame time, $t_f$, as the basic unit of time in the integration cycle, so that all the above time elements must be multiples of this. The total integration time excluding the zeroth read, $t_{int}$, is given by:
\begin{equation}
t_{int} = t_g(n-1)
\label{Eq: tint1}
\end{equation}
consistent with the definition of $t_{int}$ in \cite{Rauscher2007}.  The total integration cycle time between detector resets is given by:
\begin{equation}
t_{cycle} = t_{int} + t_{zero} + t_{dead}
\end{equation}
Assuming subtraction of the zeroth read (e.g. to correct for kTC noise), the overall efficiency (duty cycle) = $t_{int}/t_{cycle}$.  If $t_{zero} = t_{dead} = t_g$, then the efficiency = $(n-1)/(n+1)$ which is the efficiency equation given in \cite{Batalha2017}.

JexoSim also defines a simulation timestep, $t_{sim}$, which sets the time resolution for the simulation, so that $t_f$ must be an integer multiple of $t_{sim}$.  In many simulations $t_f$ can be equal to $t_{sim}$, however it may be necessary to simulate time-dependent processes at a higher rate than the detector frame rate in which case $t_{sim}$ < $t_f$.  Examples are the current pointing jitter model which runs at its own rate with time step $t_{jit}$, and the generation of `integrated' light curves that capture variation of light due to the transit within the time interval of a subintegration.

The number of groups $n$ can be user-defined. This allows $t_{int}$ to be effectively determined by the user, by entering both $n$ and $t_g$, after which $t_{int}$ obtained from Eq. \ref{Eq: tint1}. This could however possibly lead to saturation of pixels if the count within the integration exceeds the full well capacity of the pixels, $Q_{FW}$.  If any simulated pixel counts on the final stack of simulated NDR images exceed $Q_{FW}$, those pixels are flagged as saturated pixels during the data reduction pipeline in a data quality array, and thus can be identified (see Section \ref{sec 3.2}).
Alternately $n$ can be auto-calculated based on the time taken to reach a defined fraction, $\gamma$ of the full well capacity, $Q_{FW}$, on the detector pixel with the greatest count rate, $FPA(X,Y)_{max}$\footnote{In practice this is obtained by downsampling 
$cFPA(x,y)$ over a grid of points corresponding to the centre of each whole pixel and then finding the maximum value.}:
\begin{equation}
n  = \frac{[\gamma Q_{FW}/FPA(X,Y)_{max}] - t_{zero}}{t_g} +1${ }{ }{ }$\{n \in \mathbb{Z}\} 
\label{eq: n0}
\end{equation}
where $n$ is rounded down to the nearest integer value. $t_{int}$ is then given by Eq. \ref{Eq: tint1}. If all groups are equally-spaced so that $t_{zero}$ = $t_g$, this becomes:
\begin{equation}
n  = \frac{[\gamma Q_{FW}/FPA(X,Y)_{max}]}{t_g}${ }{ }{ }$\{n \in \mathbb{Z}\} 
\label{eq: n1}
\end{equation}
In Section \ref{section: HD 209458 b simulation} we simulate the transit spectrocopic observation of the hot Jupiter HD209458 b comparing JexoSim and Pandexo.  For this target observed with MIRI LRS, using $t_g$ = 0.15904 s, $\gamma$  = 1.0, and $Q_{FW}$ = 193655 e$^-$ (the default value used in Pandexo), Pandexo returned $n$ = 6. To permit a valid comparison between these simulations, JexoSim used the same values for $t_g$ and $n$ as generated by Pandexo.  However we can test if JexoSim independently returns the same value of $n$ as Pandexo. For this simulation, the maximum focal plane detector count in JexoSim, $FPA(X,Y)_{max}$ was found to be 191683 e$^-$/s. Entering this value together with the above values for $t_g$, $\gamma$ and $Q_{FW}$ into Eq. \ref{eq: n1}, returns $n$ = 6. Thus Pandexo and JexoSim both reproduce the same number of groups under the same conditions, cross-validating this aspect of each simulator.

In JexoSim we build up the NDRs from `subintegrations', where a subintegration is defined as the subarray count accumulated between two successive NDRs (i.e. the difference between them).  The exception to this is the initial (`zeroth') subintegration which is the same as the zeroth NDR. The number of subintegrations per integration cycle thus equals the number of NDRs.  In the algorithm, a `null' (zero value) array of 2-D subintegrations, $nS(X,Y,t)$, is set up in the $Timeline$ module, where $t$ is the array of time stamps  marking the end of each subintegration. These time stamps are calculated based on the timing of the corresponding integration and where each subintegration falls within the integration cycle.  Each subintegration has a duration of $t_g$, except for the zeroth subintegration which has a duration $t_{zero}$ (which is separately defined but can be set to equal $t_g$).

The user can choose how much of the observation to perform out-of-transit as a proportion of the transit duration ($T14$).
This then sets the total observing time, $t_{obs}$:
\begin{equation}
t_{obs} = N_{int}t_{cycle}$ ,{ }{ }   $ N_{int} = \frac{T14 (1 + \omega_{pre} + \omega_{post})}{t_{cycle}}${ }{ }{ }$\{N_{int} \in \mathbb{Z}\}
\label{Eq: tobs} 
\end{equation}
where $\omega_{pre}$ and $\omega_{post}$ are the fraction of $T14$ spent in pre-transit and post-transit observation respectively. $N_{int}$ is rounded down to the nearest integer, and gives the number of integration cycles in the observation. Alternately if an out-of-transit simulation is needed the user can input $N_{int}$ directly, in which case $t_{obs}$ is then calculated from Eq. \ref{Eq: tobs}.

\vspace{6pt}
 
\subsubsection{Generation of light curves} 

The $Light$ $Curve$ module controls the generation of the applicable light curves for the simulation.  This can be a primary transit, secondary eclipse or a phase curve. 
For a transit or eclipse observation, the fractional transit depth spectrum, $p(\lambda)$, is required as input. These spectra can be provided by external atmospheric codes.  A future version of JexoSim can include a dedicated atmospheric code  to generate spectra and then retrieve parameters via the JDP or alternate pipelines.
If such spectra are not available, for primary transit the user may select a `flat' fractional transit depth, fixed over all wavelengths, that would be sufficient if just bias and error bar information is needed.  Similarly an alternate input for secondary eclipse is a `smooth' planet/star emission spectrum produced by treating the planet and star as blackbodies.  
$p(\lambda)$ is rebinned to the wavelength solution for subpixel columns, $\lambda(x)$, to give $p(x)$.
 Convolution with the PSF is then simulated by first generating a stack of `null' (total area of unity) 1-D PSFs, $nPSF_{1D}(x)$. Each of these is a normalized cross-section through the maximum of the corresponding $nPSF_{2D}(x,y)$ in the $x$-axis.  These are then multiplied by the corresponding value of $p(x)$ and coadded onto a zero-value 1-D array to produce $p'(x)$, the planet-star fractional transit depth spectrum convolved with the PSF. This is then downsampled to obtain $p'(X)$.  Using $p'(X)$, we generate a 2-D array of light curves versus time, $LC(X,t)$, using the Mandel-Agol formulation \citep{Mandel2002}.  Additional inputs required for the generation of light curves are the limb-darkening coefficients, and the $z$-grid.  The latter gives the positional offsets of the planet from the centre of the stellar disc in units of $R_s$, and is related to the time grid of the observation.  
 
 JexoSim can use either `instantaneous' or `integrated' light curves.  `Instantaneous' light curves are those where we use the subintegration time base  derived in the $Timeline$ module as input to generate the z-grid.  However in reality the light curve is changing within the time period of a subintegration, thus a more accurate light curve is the `integrated' light curve which takes this into account.  For the latter we use a higher frequency time base based on $t_{sim}$ to generate the curves, which are then averaged over the duration of each subintegration, to obtain a value for each subintegration time step.  The user can judge what level of accuracy is needed, noting that the `instantaneous' light curves will generally result in a faster simulation. Best-matching limb-darkening coefficents for the stellar type are generated by the open source limb-darkening code ExoTETHyS \citep{Morello2019}\footnote{https://github.com/ucl-exoplanets/ExoTETHyS}. The current version of ExoTETHyS does not generate limb darkening coefficients above 10 \textmu m, and so the coefficients obtained for 10 \textmu m are used for all wavelengths greater than this.  
For primary transit, the light curves are normalised to unity at the out-of-transit portion.  For secondary eclipse, the light curves are normalised to unity at the mid-point of the eclipse. In the case of an out-of-transit simulation, $LC(X,t)$ will have a value of 1 throughout.

\subsubsection{Adding systematics} 
Systematic trends and distortions to the light curves can be added in the $Systematics$ module, turning $LC(X,t)$ into a modified light curve, $mLC(X,t)$, by multiplying the effects of additional time-dependent processes encoded in a 2-D matrix, $\alpha(X,t)$.  For example, the baseline light curves can have flux variations due to stellar pulsation and granulation added, as shown in \cite{Sarkar2018}.  Alternately a star spot model can add the effects of both occulted and unocculted stellar spots and faculae to the light curves.  Both these effects are the results of astrophysical models which output a version of $\alpha(X,t)$.  These effects can be applied to either out-of-transit or in-transit light curves (although the effects of occulted star spots require a full transit light curve to be used). Time-dependent instrument systematics may also be added.

\subsubsection{Pointing jitter noise} 

$nS(X,Y,t)$ is forwarded to the $Noise$ module where signal and noise are added to each subintegration. 

If pointing jitter noise is to be added, the following pathway is chosen. 
Pointing jitter power spectral density (PSD) profiles (in units of degrees$^2$/Hz vs Hz) are required that characterise the jitter in each axis, $x$ and $y$, on the focal plane.  
Timelines that contain the offsets in either the $x$ or $y$ directions due to jitter are then generated from each PSD.  Thus for each axis a jitter offset timeline is produced as follows. An array, $t'$, is produced giving the time stamp of the end of each jitter time step.  The time step for the jitter timeline, $t_{jit}$, is partially set by the requirement that the highest frequency in the model PSD be the Nyquist frequency, and thus the sampling frequency, $f_s$, must be at least twice this. However it must also satisfy the requirement that $t_f$ has to be an exact multiple of $t_{jit}$.  Thus the model PSD (for each axis) is resampled to a new frequency grid to ensure both these requirements are fulfilled, where the sampling frequency, $f_s = 1/ t_{jit}$, and with sufficient samples, $N_f$, such that the total number of samples in the corresponding time domain, $2(N_{f}-1)$, is sufficient to cover the total observing time.  For each realization, a random timeline of jitter offsets is generated for each axis as follows.  The rms amplitude of a sinusoidal Fourier component of frequency $\nu$ is given by $A_{rms}(\nu)$, which equals $\sqrt{P(\nu)}$, where $P(\nu)$ is the power of the component.  In reality, this is an $average$ power, since noisy timelines manifest due to random variations over time in both $A_{rms}(\nu)$ and the phase, $\phi(\nu)$.  Since it is algorithmically challenging to inverse Fourier transform an amplitude and phase which are both varying randomly with time, we approximate this effect by performing a single randomisation for each realization as follows.  We assume $A_{rms}(\nu)$ follows a random normal distribution $\sim N(0, {A_{rms}(\nu)}^2)$, where ${A_{rms}(\nu)}^2$ is the variance.  This distribution preserves the average power, $P(\nu)$, since the mean of the corresponding distribution of values of ${A_{rms}(\nu)}^2$ is $P(\nu)$.  We also assume the phase follows a random uniform distribution, $\sim U(0,2\pi)$.  A randomised complex number is thus obtained for each frequency component. 
This Fourier spectrum is then inverse real Fourier transformed and normalised by multiplying by $2(N_{f}-1)$.  This is used to independently generate the jitter offset timelines for the $x$ (spectral) and $y$ (spatial) directions. The offset timelines are converted from angular to spatial displacements by dividing by the plate scale, and these displacements are rounded to the nearest subpixel unit.  This gives jitter offset timelines in discrete subpixel units, $\Delta x(t')$ and  $\Delta y(t')$. 
If the rms of the jitter offsets in each axis is not Nyquist sampled at the spatial frequency of $cFPA(x,y)$, the latter grid is interpolated (upsampled by a factor $k$) 
to a higher spatial frequency to give  $cFPA(x',y')$
($x' = 0,1,2,..., 3kN_X-1$, $y' = 0,1,2,...,3kN_Y-1$), and the offset timelines are generated at this higher spatial frequency, $\Delta x'(t')$ and  $\Delta y'(t')$.
Next for each jitter timestep in $t'$, $cFPA(x',y')$ is downsampled over a 2-D grid of points corresponding to the centres of each whole pixel, offset in $x'$ and $y'$ directions by $\Delta x'(t')$ and $\Delta y'(t')$. Each point on this grid thus gives the count rate in e$^-$/s for a whole pixel over that jitter time step.  This grid is then multiplied by $t_{jit}$ to get the absolute whole pixel counts in e$^-$,  producing a jitter `subframe'. The subframes for each subintegration are coadded to form the subintegration, which is then stored in $S(X,Y,t)$.  The jitter simulation can be run with spatial jitter or spectral jitter in isolation, or with combined (spatial and spectral) jitter.

If jitter noise is not selected then the jitter algorithm is bypassed and a faster path chosen that simply downsamples $cFPA(x,y)$ over a grid corresponding to the centres of each whole pixel without any offsets. This gives the the count rate in e$^-$/s per whole pixel.  This is then multiplied by $t_{sub}(t)$, the time duration of the corresponding subintegration, to obtain the pixel counts in e$^-$ per subintegration, which is then stored in $S(X,Y,t)$. For subintegrations other than the zeroth subintegration, $t_{sub}(t)$ will equal $t_g$, and for the zeroth subintegration, $t_{sub}(t)$ will equal $t_{zero}$.

\begin{figure}
 \begin{center}
 	\includegraphics[trim={0cm 0.0cm 0cm 0cm}, clip, width=1.0\columnwidth]{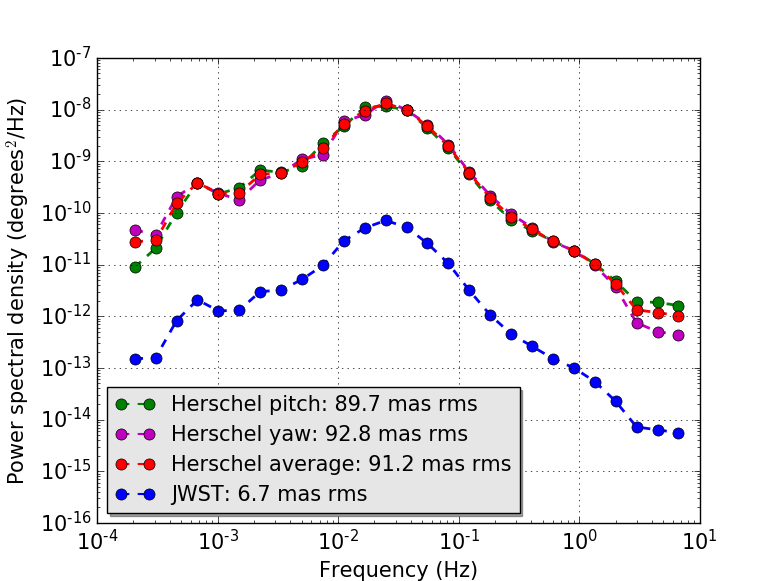}
 \end{center}

    \caption{A model pointing jitter PSD for this study was derived from a scaled version of the average Herschel PSD.}
    \label{fig:PSD}
\end{figure}
 
For JWST the 1$\sigma$ pointing stability per axis is 6.0 mas (NIRCam and NIRISS) to 6.7 mas (MIRI)  with a requirement of 16.7 mas per axis \citep{Obs2017}. Since we did not have access to a PSD for JWST, for the purposes of this paper, we use a PSD derived from the Herschel telescope pointing timeline scaled down so that it generates timelines with an rms of 6.7 mas.  We first obtain an average of the Herschel yaw ($x$-axis) and pitch ($y$-axis) PSDs, which has an rms of 91.2 mas, and then scale this down to obtain a PSD that generates a jitter timeline with rms of 6.7 mas (Figure \ref{fig:PSD}). We use this PSD for the jitter simulations in this study for both $x$- and $y$- axes.  Although the rms of the jitter timelines is set by PSD, it can be adjusted to a given value if needed by scaling the generated offsets until the desired rms is obtained.

\subsubsection{Adding the light curve and backgrounds}

At this stage, the 2-D light curve array, $mLC(X,t)$, is multiplied into the $S(X,Y,t)$. Next diffuse background sources (instrument emission and zodiacal light) are added to each pixel in each subintegration. 
The $Backgrounds$ module manages the production of the final counts resulting from diffuse backgrounds which are then applied to the pixels in the $Noise$ module.  By packaging the background models into this module, we aim to facilitate future upgrades such as potential time modulation of diffuse sources.  For the instrument emission we assume each optical surface in the OTE and instrument has an emissivity of 0.03, and emits according to its Planck blackbody spectrum.  The emission from an optical surface will be affected by the transmission of all downstream elements in the optical chain.  Since for this preliminary model we do not have the transmission profiles of every element individually (e.g. transmission files for `internal optics' will have the combined transmission of many elements) we approximate this effect by taking the final combined transmission for the OTE, $\eta_{tel}$, and that for instrument channel, $\eta_{ch}$ and divide the transmission equally among each of the optical elements.  Thus if there are $N_{tel}$ optical surfaces in the OTE, and $N_{ch}$ optical surfaces in the instrument, the transmission for each individual optical surface, $\eta_{tel}'$ and $\eta_{ch}'$ is calculated as follows:
\begin{equation}
\eta_{tel}' = {\eta_{tel}}^{1/N_{tel}}$, { }{ }{ }$ \eta_{ch}' = {\eta_{ch}}^{1/N_{ch}}
\end{equation}
Then the total emission at the end of the OTE optical chain is given by:
\begin{equation}
\varepsilon(T,\lambda)_{tel} = \sum_{i=1}^{i=N_{tel}} 0.03B_{\lambda}(T){\eta_{tel}'}^{N_{tel}-i} 
\end{equation}
The final emission at the end of the instrument optical chain is given by: 
\begin{equation}
\varepsilon(T,\lambda)_{ch} = \varepsilon(T,\lambda)_{tel}{\eta_{ch}} + \sum_{i=1}^{i=N_{ch}} 0.03B_{\lambda}(T){\eta_{ch}'}^{N_{ch}-i} 
\end{equation}
where $B_{\lambda}(T)$ is the Planck function, and $T$ is set to 50 K for the OTE, and 40 K for the instrument channels with the exception of MIRI where is it set to 7 K.  We have assumed $N_{tel}=4$. For the instruments we currently estimate $N_{ch}$ from publicly available diagrams of the optical paths. 
As a result of the conservation of entendue, the count in e$^-$/s per whole pixel $X(\lambda)$ on each pixel row is given by:
\begin{equation}
C_{\varepsilon}(X) = \left[ \varepsilon(T,X)_{ch}A_{pix}\Omega_{pix}QE(X)\frac{hc}{\lambda(X)}\Delta \lambda(X) \right] \circledast W_s(X)
\end{equation}
where $A_{pix}$ is the area of the pixel, $\Omega_{pix}$ is the solid angle subtended at the pixel and $\Delta \lambda(X)$ is the wavelength span across a pixel.  The wavelength $\lambda$ assigned to pixel $X$ is determined by the wavelength solution for a point source.  For diffuse radiation however, rays of the same wavelength will be coming in from a range of angles, this range being restricted by the slit.  This results in the wavelength solution (and therefore the spectrum on the $X$-axis) being shifted and repeated on the detector over a distance range that equals the projection of the slit onto the detector. This means the count on a given pixel is the result of co-adding counts from the repeated spectra.  We simulate this by convolving the count rate per pixel with a filter function, $W_s(X)$, which is a box function of height unity and length equal to the projected slit width on the detector. For NIRSpec, we use a slit width of 16 pixels. For slitless modes, the range of angles is no longer restricted and the spectrum repeats across the entire detector, with all $X$ pixels receiving all wavelengths within the bandpass of the channel transmission, and thus all having the same count. We simulate this by making the length of $W_{s}(X)$ equal to twice the number of $X$ pixels.  More precisely in the presence of a slit, some emissions will arise from optical surfaces that go through the slit and some from surfaces that do not, and these should be managed separately, but this detail has not yet been implemented.  We assume no $Y$ dependency on the diffuse signal, and thus $C_{\varepsilon}(X)$ is applicable to all rows on the detector array. $C_{\varepsilon}(X)$ is multiplied by $t_{sub}(t)$ to get the counts in e$^-$ for each subintegration in $S(X,Y,t)$.

\begin{figure}
 \begin{center}
 	\includegraphics[trim={0cm 0.0cm 0cm 0cm}, clip, width=1.0\columnwidth]{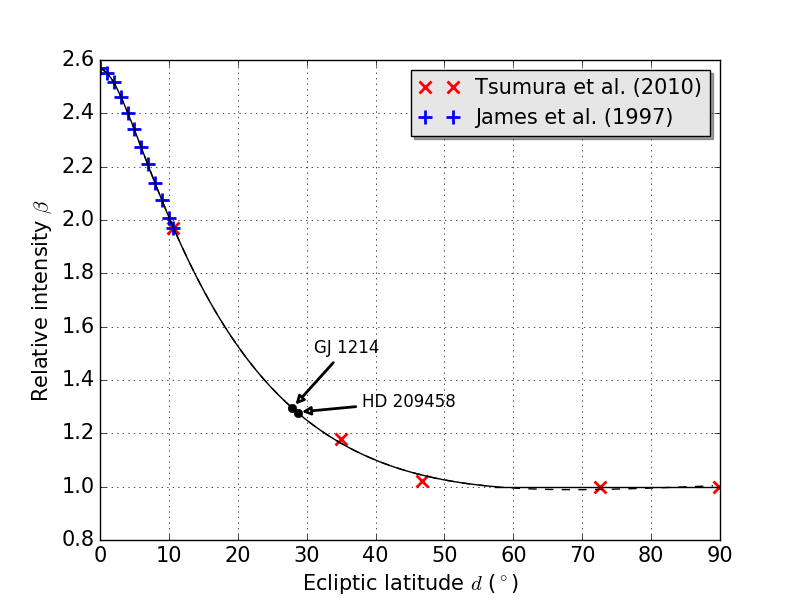}
 \end{center}
 
    \caption{Relative intensity, $\beta$, of zodiacal light with ecliptic latitude, $d$. Data points are explained in the text. The solid black line shows the model used in JexoSim, which uses the 7th order polynomial from Equation \ref{Eq: zodi trend} to find $\beta$ for $0>d\leq 57.355 ^\circ$.  Above 57.355$^\circ$ we assume $\beta$ = 1.0 since the polynomial fit (dashed black line) falls slightly below 1.}
    \label{fig:zodi}
\end{figure}

For zodiacal dust cloud light we use a formula \citep{Pascale2015} that models spectral brightness in W/m$^2$/\textmu m/sr with reflective and emissive components:
\begin{equation}
I_{zodi}(\lambda) = \beta \left[ 3.5 \times 10^{-14}B_{\lambda}(5500K) + 3.58 \times 10^{-8}B_{\lambda}(270K) \right]
\end{equation}
When the coefficient $\beta=1$, the formula approximates the intensity at the South Ecliptic Pole as shown in \cite{Leinert1998}. For sources near to the ecliptic where the dust density is increased the coefficient may be increased.  Figure \ref{fig:zodi} shows how the $\beta$ factor may increase with decreasing latitude. Here we took data points from two studies which gave relative intensities of the zodiacal light at different ecliptic latitudes. \cite{Tsumura2010} use the Low Resolution Spectrometer onboard the Cosmic Infrared Background Experiment (CIBER) to obtain five data points from 
90$^\circ$ (i.e the ecliptic pole) to 10.6$^\circ$.  We normalize all points to the ecliptic pole data point to find the relative intensity of each.  Next we take we take the average of six intensity curves from \cite{James1997} sampled at points between 0 to 10.6$^\circ$.  The intensity of these points relative to the 10.6$^\circ$ data point is found.  These values are then multiplied by the relative intensity of the 10.6$^\circ$ data point from \cite{Tsumura2010} to give intensity values relative to the ecliptic pole, $\beta$. The results are shown in Figure \ref{fig:zodi}.  If $d$ is the ecliptic latitude in degrees, and $\zeta = \log_{10}(d+1)$, we find that a good fit can be obtained to these data points using the following polynomial equation:
\begin{equation}
\begin{split}
\beta = 
-0.22968868 \zeta^7 + 1.12162927 \zeta^6 - 1.72338015\zeta^5 \\
+ 1.13119022 \zeta^4 - 0.95684987\zeta^3+ 0.2199208\zeta^2\\
- 0.05989941\zeta  +2.57035947\\
\end{split}
\label{Eq: zodi trend}
\end{equation}
Since the polynomial falls slightly below 1 at $d>$ 57.355$^\circ$ (dashed black line in Figure \ref{fig:zodi}), we use the above polynomial to find $\beta$ from 0$^\circ$ to 57.355$^\circ$, and use a value of 1 above 57.355$^\circ$ (solid black line in in Figure \ref{fig:zodi}). 
This relationship is an approximation since the interplanetary dust cloud is non-uniform,  however we utilise this model in this paper to find $\beta$ values for the exosystems GJ 1214 and HD 209458 which are simulated for this paper (Figure \ref{fig:zodi}).
The count in e$^-$/s per pixel $X(\lambda)$ on each pixel row is given by:
\begin{equation}
C_{zodi}(X) = \left[ I_{zodi}(X)A_{pix}\Omega_{pix}QE(X)\frac{hc}{\lambda(X)}\Delta \lambda(X)  \right]  \circledast W_{s}(X)
\end{equation}
Convolution with $W_s(X)$ is again performed for the same reasons described above. $C_{zodi}(X)$ is finally multiplied by $t_{sub}(t)$ to get the counts in e$^-$ for each subintegration in $S(X,Y,t)$.

\subsubsection{PRNU}
\label{sec:PRNU}

The photo-response non-uniformity (PRNU) is the inter-pixel QE variation.  In JexoSim this is applied in the $Noise$ module as a 2-D array, $PRNU(X,Y)$, to all subintegrations after the generation of addition of backgrounds. The PRNU grid encodes the rms variation around the baseline QE, $QE(X)$ which was applied at an earlier step. A pre-defined grid of values can be used or a random grid generated based on an rms value defined by the user, e.g. $\pm$3\% \citep{Ressler2015} which we adopt for this study for all detectors. This grid can then be used in data reduction as the `flat field'. An uncertainty can be added to the PRNU knowledge by the user, e.g $\pm$0.5\% \citep{Pirzkal2011}, which we also adopt in this study. This uncertainty is not flat-fielded out. 
To incorporate intra-pixel response functions which produce inter-pixel cross-talk (e.g. the bell-shaped function from Figure \ref{IPRF} B), $PRNU(X,Y)$ is first interpolated to the same spatial frequency as the Nyquist-sampled IPRF.
The two function are then convolved, and the convolved function is then downsampled at the whole pixel positions to reconstitute $PRNU(X,Y)$. This latter grid has an adjusted QE variation for each pixel that includes the effect of diffused charge from neighbouring pixels which have different QE. This way $PRNU(X,Y)$ can be used to apply the effects of both interpixel QE variations and crosstalk in one step.
Additional detector systematics that are planned for JexoSim  include non-linearity, cosmic ray hits, persistence, and bad/hot pixels. Adding effects such to JexoSim will enable testing of corresponding data reduction steps.

\subsubsection{Final steps: dark current, Poisson noise, formation of NDRs and read out noise}

Next, dark signal is added to each pixel calculated from the dark current, $I_{dc}$, multiplied by $t_{sub}(t)$.  $I_{dc}$ values for NIRISS, NIRCam and NIRSpec were obtained from the JWST user documentation \citep{NIRISS2019,NIRCam2019,NIRSpec2019} and for MIRI from \cite{Rieke2015b} (Table \ref{table:Config}).
 In future versions, dark current variations across the detector array can be included, which would in turn require the use of dark images in data reduction. 
 Next, random variations are applied to each pixel in each subintegration to simulate Poisson `shot' noise. If photon noise from the target star (more precisely the target exosystem) is required in isolation, then the other Poisson noise sources (i.e. the diffuse backgrounds and dark current) can be turned off. The same method can be applied to obtain the Poission noise contribution of any of the backgrounds or dark current in isolation. Next, the subintegrations within each integration cycle, $S_i$ ($i=0,1,2,...,n-1$), are coadded in sequence to generate the NDRs for each exposure, $N_j$ ($j=0,1,...,n-1$). $N_j$ is thus given by:

\begin{equation}
N_j = \sum_{i=0}^{i=j}{S_i} 
\end{equation}
 
\noindent This way an array of 2-D NDRs versus time, $N(X,Y,t)$, is generated. Finally variations in the signal due to read out noise are randomly generated for each pixel assuming a Gaussian distribution with standard deviation, $\sigma_{ro}$, the read out noise for one read. We use the same sources for $\sigma_{ro}$ as for $I_{dc}$.  For MIRI, \cite{Rieke2015b} give a value of 14 e$^-$ based on Fowler-8 sampling. Since $\sigma_{Fowler-8} = \sqrt{2}\sigma_{ro}/\sqrt{8}$ we use a value of 28 e$^-$ for $\sigma_{ro}$. The variations are added to each pixel in each NDR to give the final signal and noise array, $N'(X,Y,t)$.
This array can either be packaged into FITS file format to resemble a JWST data level product for input into external pipelines, or can be passed into the associated pipeline, the JexoSim Default Pipeline where it is processed to provide immediate results for the user.

\subsection{JexoSim Default Pipeline and analysis options}
\label{sec 3.2}

The associated JexoSim Default Pipeline (JDP) is a basic data reduction pipeline that processes the image stack, $N'(X,Y,t)$, and produces the a number of different result products depending on the choice of the user. A data quality array is also set up used that tracks each pixel on each NDR and then the integration image. This can be used to flag known inoperable pixels, saturated pixels or bad ramps. The image stack is processed with the following steps:
1) dark subtraction,
2) flat fielding,
3) background subtraction, 
4) processing the ramp: correlated double sampling (CDS) (which is the same as last-minus-first or LMF processing), up-the-ramp slope fitting (UTR), or Fowler sampling (also known as multiple CDS or MCDS),
5) decorrelation of pointing jitter (applied only if simulations include jitter),
6) aperture masking,
7) extraction of 1-D spectra,
8) binning of the 1-D spectra into spectral resolution element sized bins (`$R$-binning') or fixed bins of a given number of pixel columns in width,
9) fitting model light curves and extraction of the planet spectrum.
Aperture masking is achieved by subpixelising the $Y$-axis, and fitting an aperture over the signal. The 
width of the aperture in the spatial direction can be varied, and can be a function of the wavelength, e.g. the Airy disc diameter, 2.44 $F\lambda$ (where $F$ is the final f-number or focal ratio for the channel, and $\lambda$ is the wavelength per pixel column).
For simulations that include pointing jitter, a moving aperture (in the $Y$-axis) is applied per image, where the aperture is re-centered on the peak of the signal in the $Y$-axis for each image.  Pointing jitter decorrelation is achieved in two stages: 1) elucidating the $X$ (spectral) and $Y$ (spatial) offsets of each integration image relative to a reference image, and 2) shifting the images by the calculated offsets to reverse the effect of jitter movements.  The former step can be achieved using either the pointing timeline information or by image cross-correlation. The latter step can be achieved either through 2-D cubic interpolation or by applying 2-D Fourier transform to the images to apply phase shifts (i.e. reverse the offsets) in Fourier space. For this paper, we use a combination of pointing timeline information and the 2-D interpolation method.  This step may be in addition to what official JWST pipelines consider, however we have retained it in this study for completeness since we do simulate jitter for some simulations, but it can be omitted if needed.  For $R$-binning, each 1-D spectrum (which is pixelised), is divided into spectral resolution element-sized bins. The boundaries of these bins will generally cross whole pixels. Whole pixels falling entirely within a bin contribute all their signal to that bin.  The boundary pixels are subdivided, with a proportion of their signal given to each bin. The proportion given to each bin is weighted according to the gradient of the signal over the boundary pixel and where the division occurs along the length of the pixel.  This method  results in smoothly sized resolution element bins.  

As JexoSim is further developed and additional systematics are added into the simulation, so additional corresponding steps in the JDP may be needed. Alternately official JWST pipelines can be run using the JexoSim output. Various results can be obtained through the JDP and some options are given below.

\subsubsection{Out-of-transit signal and noise performance estimates} A simple but informative performance measurement is to find the standard deviation, $\sigma_s(\lambda)$, of the out-of-transit signal in each spectral bin, $s(\lambda,t)$. Here $s(\lambda,t)$ is the signal per spectral bin per integration after processing through the pipeline.  JexoSim can perform out-of-transit simulations for this purpose.  In this case the JDP stops at step 8. This can be used to find the signal-to-noise ratio (SNR) per bin  or the noise variance per unit time.  The latter quantity however is constant with time only for uncorrelated noise. By repeating the simulation with individual noise sources turned on and all other sources off, a `noise budget' can be built up showing the contribution of each noise source to the overall noise.  This can help to optimize signal-to-noise  by developing observing strategies and  focussing on those data processing steps that help to mitigate the most significant sources of noise.  It can also be used to show compliance with mission requirements on noise.  If $N_{int}$ is the total number of integrations, of which half are in-transit and half are out-of-transit, then the uncertainty on the final transit depth spectrum (i.e. the `error bars'), $\sigma_p(\lambda), $ can be estimated as:
\begin{equation}
\sigma_p(\lambda) = \frac{2}{ \sqrt{N_{int}}} \frac{\sigma_s(\lambda)}{ \left\langle s(\lambda,t) \right\rangle } 
\end{equation}

This estimate assumes noise which is uncorrelated in time, and does not take into account the curvature of a light curve due to limb-darkening.

\subsubsection{Allan deviation plots and estimates for $\sigma_p$} 
\label{sec: Allan}
To account for the fact that time-correlated noise may integrate down more slowly than uncorrelated noise, which in turn will affect the relative contribution of the different noise sources in the overall budget, an Allan deviation plot can be produced.  Here, for each spectral bin, the timeline of final integration signals, $s(\lambda,t)$, is divided into contiguous segments, where each segment contains a number of integration cycles, each of duration $t_{cycle}$.  If the total time length of the segment is $\tau$, and the number of integration cycles in the segment is  $N_{\tau}$, then $\tau = N_{\tau} t_{cycle}$. $\tau$ is increased progressively (by increasing $N_{\tau}$), and for each value of $\tau$,
the mean signal of each segment, $\left\langle s \right\rangle $, is found.  The standard deviation of the mean signals is then obtained,  $\sigma_{\left\langle s \right\rangle }$.
This is effectively directly measuring the `standard error on the mean'.  
The fractional noise, $\sigma_{\left\langle s \right\rangle}   /  \mu_{ \left\langle s \right\rangle }$, is then obtained, where $\mu_{ \left\langle s \right\rangle }$ is the mean of all values of  $\left\langle s \right\rangle$  in the timeline.
The Allan deviation chart plots the fractional noise versus $\tau$. 
 For uncorrelated noise, the fractional noise should fall with a power law exponent of -0.5, but this may not be the case for correlated noise.  Thus depending on the value of $\tau$, the relative contributions of correlated and uncorrelated noise may vary. The fractional noise value of most interest will be at the time scale of the transit, i.e. when $\tau=T14$, and thus noise budgets that find the fractional noise contributions of different noise sources at such a time scale will provide the best representation of the relative impact of each noise source on the mean signal.
If total noise is considered, the fractional noise at $T14$ can be used to estimate the noise on the transit depth $p(\lambda)$.  If an observation has an equal amount of time out-of-transit as in-transit:
\begin{equation}
\sigma_p(\lambda) = \sqrt{2}\frac{   \sigma_{   \left\langle s \right\rangle  } }{ \mu_{ \left\langle s \right\rangle}} (\lambda)    
\label{Eq: sigma_p}
\end{equation}
where $\sigma_{\left\langle s \right\rangle}   /  \mu_{ \left\langle s \right\rangle }$ is the fractional noise at $T14$.
This formula assumes no correlation between the in- and out-of-transit data and so might underestimate the noise for correlated noise (although less so than if the Allan deviation approach was not used at all). Another method that accounts for the correlation between in- and out-of-transit data for estimating $\sigma_p(\lambda)$ using an out-of-transit timeline is described in detail in \cite{Sarkar2018}.
Due to computational limitations it may not be possible 
to generate a sufficiently long timeline
to find the fractional noise at $T14$ directly if at least 20 segments are needed to find a standard deviation, since the data become more scattered as the number of bins falls (as $\tau$ increases).  However, the Allan deviation plot allows the power law trend of  $\sigma_{\left\langle s \right\rangle}   /  \mu_{ \left\langle s \right\rangle }$ vs $\tau$ to be estimated (by fitting a straight line in log-log space) and this can then be   extrapolated to find the fractional noise at $\tau=T14$.  We adopt this method for the noise budgets shown in this paper.  It may be that for correlated noise the power law does not stabilize until $\tau$ reaches larger values, in which case the straight line should be fitted to the later points. In this paper we fit lines to the latter 75\% of points.

\begin{figure*}
 \begin{center}
 	\includegraphics[trim={1cm 0cm 1cm 1cm}, clip, width=1.8\columnwidth]{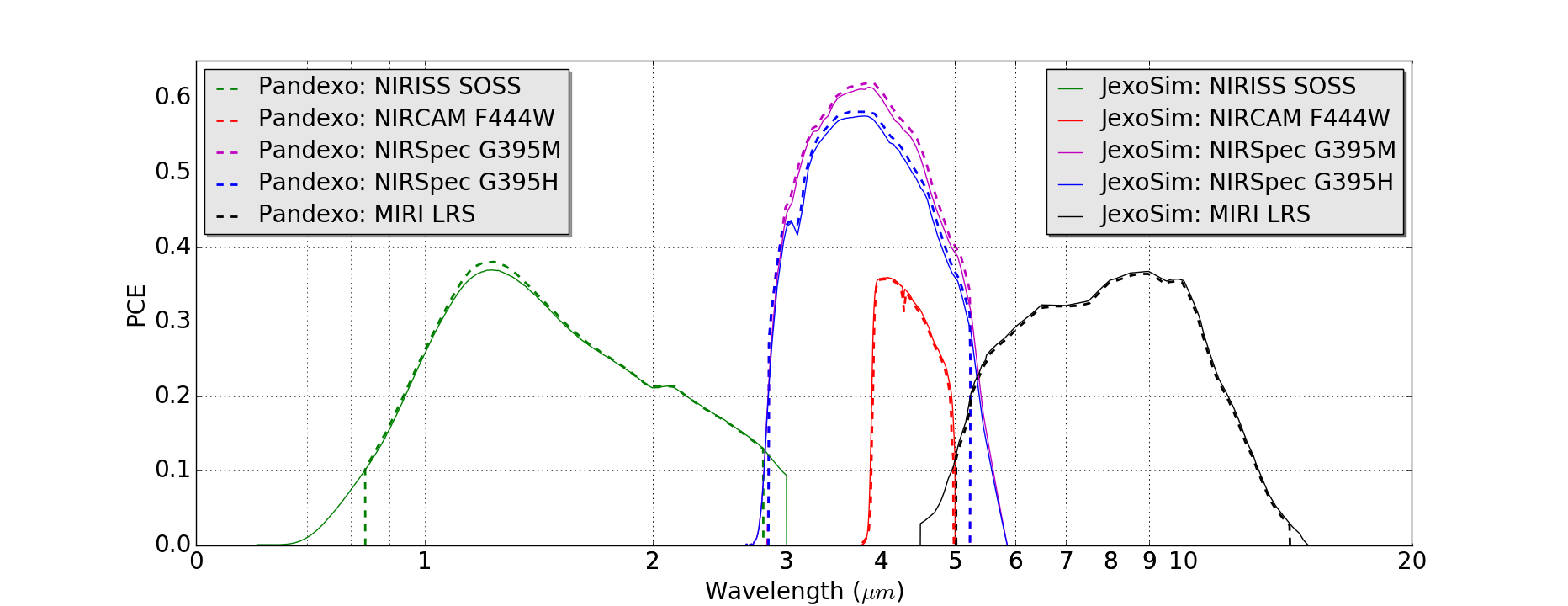}
 \end{center}
\vspace{-0.3em}  
    \caption{Comparison between over photoconversion efficiency (PCE) profiles for JWST channels used in JexoSim and Pandexo.}
    \label{fig:PCE}
\end{figure*}

\subsubsection{Planet spectrum with error bars} 
JexoSim in full transit simulation mode followed by the JDP can generate simulated exoplanet spectra binned to a user-defined spectral resolving power, $R$ (where $R$ = $\lambda/\Delta \lambda$). This mimics the expected final result of a real transit observation after data pipeline processing, giving the extracted spectrum of fractional transit depths, $p(\lambda)$, with errors bars showing the 1$\sigma$ uncertainty, $\sigma_p(\lambda)$.  Such spectra could then be used as input for spectral retrieval algorithms that would then return the maximum likelihood solution for the atmospheric composition and structure.  This is probably the most rigorous approach to assess specific `science challenges'.  Such challenges may involve questions such as: can the the C/O ratio be successfully extracted for a given planet, and if so how many transit or eclipse observations will be needed? Another challenge might be the question of whether the contaminating effects of star spots and faculae on the spectrum for a given exosystem significantly affect the accuracy and precision of  
$p(\lambda)$.  The user can utilize their own light curve fitting algorithm if they wish by taking the data light curves generated at step 8 of the JDP.  Alternately the JDP will process the light curves to extract the transit depth using its own default algorithm.  By default we use scipy.optimize.minimize with the Nelder-Mead simplex algorithm, minimizing a chi-squared function, to fit a model Mandel-Agol light curve to each spectral bin to retrieve a transit (or eclipse) depth.  The error bar can be obtained through: 1) an out-of-transit simulation adopting one of the methods described above to estimate $\sigma_p(\lambda)$, or 2) using a multi-realization Monte Carlo simulation.  The latter has the benefit that through repeated realizations, a distribution of recovered transit depths can be obtained.  The standard deviation of this distribution effectively gives a direct measure of the probability distribution on the transit depth and should capture the effect of any non-Gaussian noise sources as well as time-dependent correlated noise.  The mean of such a distribution may also reveal any residual wavelength-dependent systematic biases when compared to the input planet spectrum (which is known), as could occur due to star spot contamination or an instrumental systematic process.

\section{JexoSim Photon Conversion Efficiency}

The first stage in validating JexoSim is to compare our final model photon conversion effiency (PCE) profiles with those publically available for Pandexo for five instrument modes (Figure \ref{fig:PCE}).  The PCE gives the fraction of incident photons per unit wavelength that are available for conversion to electrons after factoring in all the OTE and instrument channel transmissions as well as the wavelength-dependent quantum efficiency of the detector.  This gives a measure of how well the transmissions have been captured in the current version of JexoSim used for this paper compared to those in Pandexo.
For Pandexo we estimate the PCE from the publicly available PCE charts from the Pandexo homepage\footnote{https://exoctk.stsci.edu/pandexo/calculation/new}. We can see that overall the PCE profiles used in JexoSim  match those of Pandexo quite well, indicating JexoSim is capturing the transmissions and quantum efficiency to a similar level of accuracy. There are some differences in the cutoff wavelengths used in Pandexo at the band edges, compared those in the PCE profiles derived for JexoSim, but these will have no consequence to the simulated results in this paper as they are outside of the designated wavelength bands for the instruments. The other small discrepancies seen may be due to errors in extracting the Pandexo profiles or possibly small transmission differences not currently accounted for in JexoSim.  

\begin{table*}
\begin{center}
\caption{Instrument channel configurations used in JexoSim for this paper.  For MIRI, we obtain all parameters except the plate scale from \protect\citep{Ressler2015}.  The MIRI read noise is derived from a quoted value of 14 e$^-$ for Fowler-8 sampling.
For NIRSpec, the read noise is obtained from \protect\cite{Giardino2012}. 
Remaining parameters were obtained from the Cycle 1 user documentation for each instrument \protect\citep{NIRISS2019, NIRCam2019, NIRSpec2019, MIRI2019}. For NIRSpec, the dark current, $I_{dc}$, and full well capacity, $Q_{FW}$, are the average of the two detector values given in the user documentation.
$\lambda$ is wavelength range, $\Delta_{pix}$ is pixel length, $\phi_x$ and $\phi_y$ are the plate scales in $x$ and $y$ directions respectively, and $T$ is the operating temperature of the channel.  Other abbreviations are explained in the text. }
\label{table:Config}

\setlength{\tabcolsep}{12pt}
\begin{tabular}{lcccc} 
\hline
Instrument&NIRISS&NIRCam&NIRSpec&MIRI\\
mode&SOSS&Grism-R/F444W&G395M/F290LP &LRS \\
&(order 1)&(long-wave channel)&BOTS &  \\
\hline
$\lambda$ (\textmu m)&
0.85-2.8&
3.9-5.0&
2.87-5.1&
5-12\\

Subarray mode&SUBSTRIP256&
SUBGRISM64&
SUB2048&
SLITLESSPRISM\\

Subarray size&
256 x 2048&
64 x 2048&
32 x 2048&
72 x 416\\

$W_s$ ($\Delta_{pix}$)&
N/A&
N/A&
16&
N/A\\

$\sigma_{ro}$(e$^-$)&
12.95&
9.55&
12&
28\\

$I_{dc}$(e$^-$/s)&
0.0257&
0.027&
0.0075&
0.2\\

$\Delta_{pix}$ (\textmu m)&
18 &
18 &
18 &
25\\

$Q_{FW}$ (e$^-$)&
100k &
83.3k &
57.75k &
250k \\

$\phi_x$ $(^\circ \times 10^{-5}/\Delta_{pix})$&
1.817 &
1.750 &
2.777 &
3.056 \\
$\phi_y$ $(^\circ \times 10^{-5}/\Delta_{pix})$&
1.828 &
1.750 &
2.777 &
3.056 \\
$T$ (K)&
40 &
40 &
40 &
7 \\

\hline
\end{tabular}
\end{center}
\end{table*}

\section{Focal plane and photon noise simulations}
\label{section: fp sims}
Next, we validate JexoSim noise results against those from Pandexo v1.0 as presented in \cite{Batalha2017} (hereafter B17) as well as results from independent instrument simulators presented in the same paper (listed in Section \ref{Intro}). To permit a valid comparison we simulate the same host star as in B17: surface temperature 4000 K, log $g=4.5$ and Fe/H = 0, normalised to J magnitude of 8, for which we utilise the corresponding PHOENIX model spectrum.  The JexoSim model configurations of the 4 instrument modes tested are given in Table \ref{table:Config}. We elect to compare only the exosystem photon noise results using LMF mode.  This gives the simplest comparison test, and is not subject to differences in background models, dark current or read noise values used between the different simulators. It also excludes noise sources such as jitter noise which are simulated in JexoSim but not in Pandexo.  If there is reasonable agreement in the photon noise, it gives confidence in the fundamental radiometric model used for JexoSim, as well as key steps in the pipeline, such as $R$-binning.  JexoSim was therefore run with all other noise sources and backgrounds switched off and no PRNU applied.  For these tests JexoSim was run in out-of-transit mode, with $N_{int}$ set to 5000 in to sample the standard deviation of the signal. $t_{int}$ was set individually for each channel to match the times used in B17 as described further below. The JDP performed correlated double sampling (CDS), i.e. last-minus-first (LMF), processing (step 4) and omitted all other steps except steps 7 and 8. Since LMF is being used, for computation efficiency, the simulations were run with $n=2$, and $t_g$ selected to produce the required value of $t_{int}$. The binned $R$ power used for each instrument mode was estimated to match those from B17. This was performed through a measurement of the wavelengths of data points in noise vs wavelength charts presented in B17. We first find the wavelength of each point $i$, $\lambda[i]$. Then we define $R[i] = 0.5(\lambda[i]+\lambda[i+1]) / (\lambda[i+1]-\lambda[i])$, and use the average value of $R[i]$ over all the bins as $R$ in each case. This is different to the way $R$ was defined in B17 (Batalha, N. E., personal communication) where $R = \lambda[i]/(\lambda[i+1]-\lambda[i-1])$.

\subsection{NIRISS}

\begin{figure}
 \begin{center}
 	\includegraphics[trim={1cm 0cm 2cm 0cm}, clip, width=1.0\columnwidth]{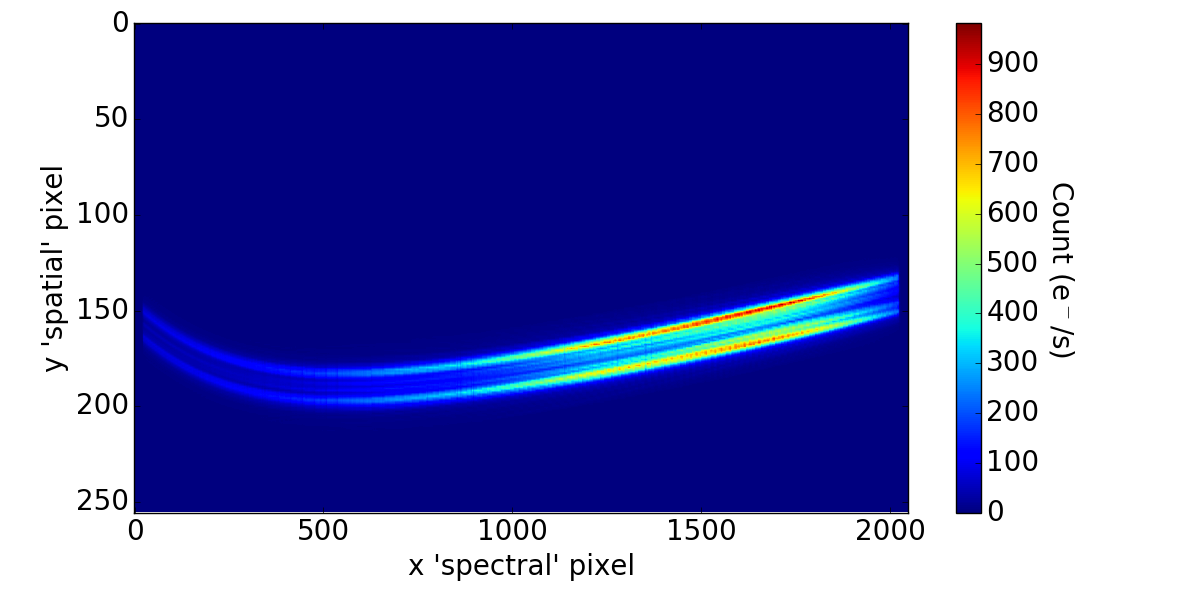}
 \end{center} 
    \caption{NIRISS SOSS focal plane image simulated in JexoSim}
    \label{fig:NIRISS-fp}
\end{figure}

\begin{figure}
 \begin{center}
 	\includegraphics[width=1.0\columnwidth]{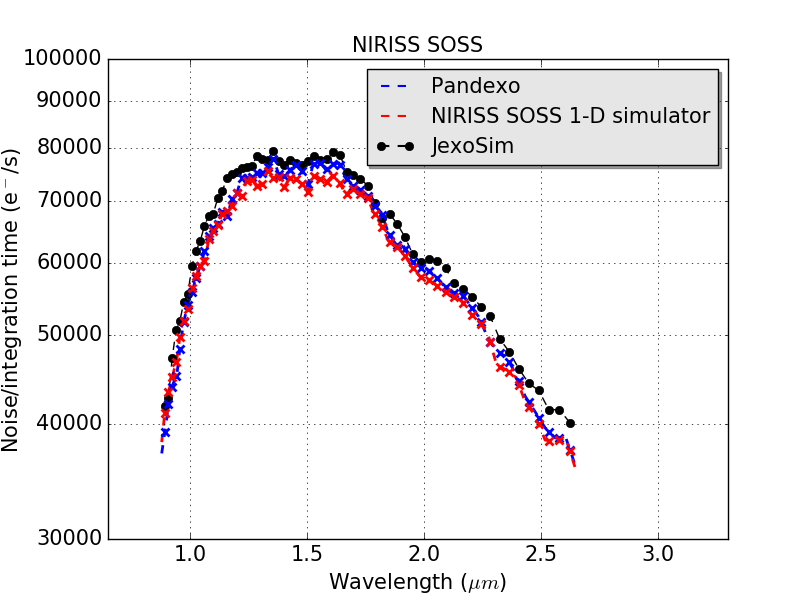}
 \end{center}
    \caption{NIRISS SOSS out-of-transit photon noise per unit integration time. JexoSim is compared compared to Pandexo and NIRISS SOSS 1-D simulator results from B17. Binned $R$ power in JexoSim is 58.}
    \label{fig:NIRISS}
\end{figure}

The Near Infrared Imager and Slitless Spectrograph (NIRISS) in Single Object Slitless Spectroscopy (SOSS) mode is enabled by using the GR700XD grism from the pupil wheel. This grism has an intrinsic $R$ power of $\sim$ 700 at 1.25 \textmu m for the 1st order spectrum \citep{NIRISS2019}, and covers a wavelength range from 0.85-2.8 \textmu m.  Two additional weaker orders are also generated.  Unique among the instruments simulated, the NIRISS SOSS mode produces a significantly curved spectral trace on the focal plane.  In addition, the PSF is elongated in the cross-dispersion direction by a cylindrical lens as discussed above. For this paper we simulated only the 1st order spectrum using the SUBSTRIP256 subarray.  The focal plane image from JexoSim is shown in Figure \ref{fig:NIRISS-fp}.  Replicating the observational parameters described in B17, $t_{int}$ was set to 2 hours. The 1-D spectra were binned to an $R$ power of 58, based on the estimated measurement of the distribution of points from the chart of noise vs wavelength for NIRISS in B17, as explained above. To match the format of results from B17, $\sigma_{s}(\lambda)$ (i.e. the standard deviation of the signal per spectral bin) was divided by $t_{int}$ to obtain the noise per unit integration time.  The results are shown in Figure \ref{fig:NIRISS}, together with results for photon noise from Pandexo and the NIRISS SOSS 1-D simulator estimated from B17.  The Pandexo and NIRISS SOSS 1-D simulator lines are interpolated to the JexoSim spectral bin wavelength points (crosses on Figure \ref{fig:NIRISS}).  Comparing these points, we find that the mean percentage difference (and standard deviation) of JexoSim from Pandexo is +4.1 $\pm$ 2.4 \%, and from the NIRISS SOSS 1-D simulator results it is +5.4  $\pm$ 1.9 \%.  Over all comparison points, JexoSim is always within 12.3 \% of Pandexo, and within 9.2 \% of the NIRISS SOSS 1-D simulator.  Since JexoSim produces noise in a stochastic simulation, we would expect variations around the trend in noise from bin to bin, but since the mean deviations from the other simulator results are approximately within 5 \%, we consider this a reasonably good agreement with the other simulator results.  This would validate the accuracy of the fundamental radiometric model used in JexoSim.  The inconsistencies that do exist may be due to several factors: small differences in the simulations, e.g. the exact PHOENIX spectrum used or how this was the normalised, inaccuracy in the estimation of the binned $R$ power or differences in the $R$-binning methods used, small differences in the transmissions used or in the collecting area of the primary mirror.

\begin{figure}
 \begin{center}
 	\includegraphics[trim={1cm 0cm 2cm 0cm}, clip, width=1.0\columnwidth]{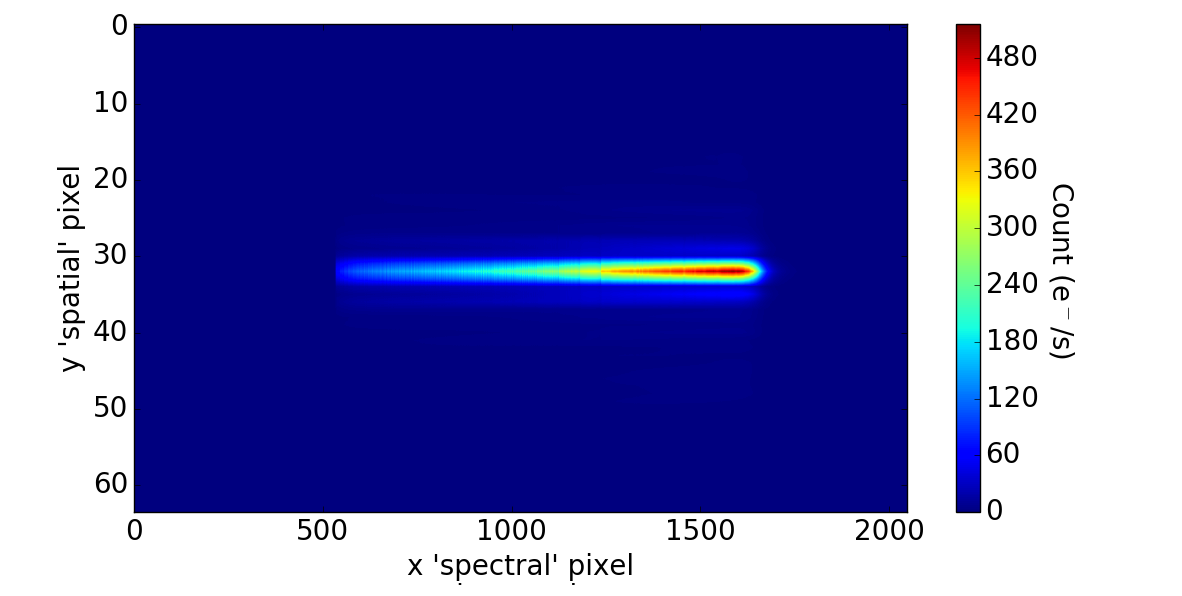}
 \end{center} 
    \caption{NIRCam Grism-R/F444W focal plane image  simulated in JexoSim}
    \label{fig:NIRCAM-fp}
\vspace{-2.2mm}
\end{figure}

\begin{figure}
 \begin{center}
 	\includegraphics[width=1.0\columnwidth]{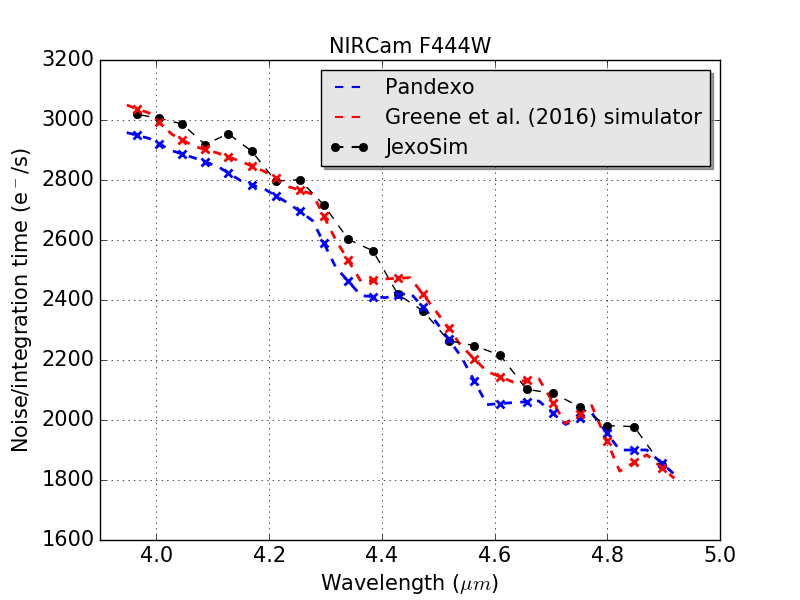}
 \end{center}
    \caption{NIRCam Grism-R/F444W out-of-transit photon noise per unit integration time. JexoSim is compared compared to Pandexo and \protect\cite{Greene2016} simulator results from B17. Binned $R$ power in JexoSim is 100.}
    \label{fig:NIRCAM}
\end{figure}

\subsection{NIRCam}
The NIRCam grism time series observation mode can be used for transit spectroscopy. NIRCam has short and long wave channels split using a dichroic. There are two grisms, both operated through the long wave channel, both with an intrinsic $R$ power of $\sim$ 1600 and wavelength range of 2.4-5.0 \textmu m \citep{NIRCam2019}; one grism disperses spectra along columns and the other along rows.  The grisms can be combined with different wide field filters.  For this paper we simulate the F444W filter (3.9-5.0 \textmu m) and slitless time series spectroscopy with the SUBGRISM64 subarray. We assume use of Grism-R, but the model does not change for the alternate grism.
The focal plane array image generated in JexoSim is shown in Figure \ref{fig:NIRCAM-fp}. Matching the overall integration time used in B17, we use a $t_{int}$ for each LMF integration of  of 18.36 s.
This time the spectra are binned to an $R$ power of 100, again based on the measured distribution of the data points in B17. The noise per unit integration time for each spectral bin was again obtained over 5000 integrations and the results are shown in Figure \ref{fig:NIRCAM}. We find that the JexoSim results have an average difference (and standard deviation) of
+3.0 $\pm$ 2.2 \% from Pandexo, and +1.2 $\pm$ 2.1 \% from the \cite{Greene2016} NIRCam simulator result.  Over all points compared, JexoSim is always within 7.8 \% of Pandexo and within 6.4\% of the \cite{Greene2016} simulator.  We consider these results to be in reasonably good agreement, again validating JexoSim and key steps in the pipeline.

\begin{figure}
 \begin{center}
 	\includegraphics[trim={1cm 0cm 2cm 0cm}, clip, width=1.0\columnwidth]{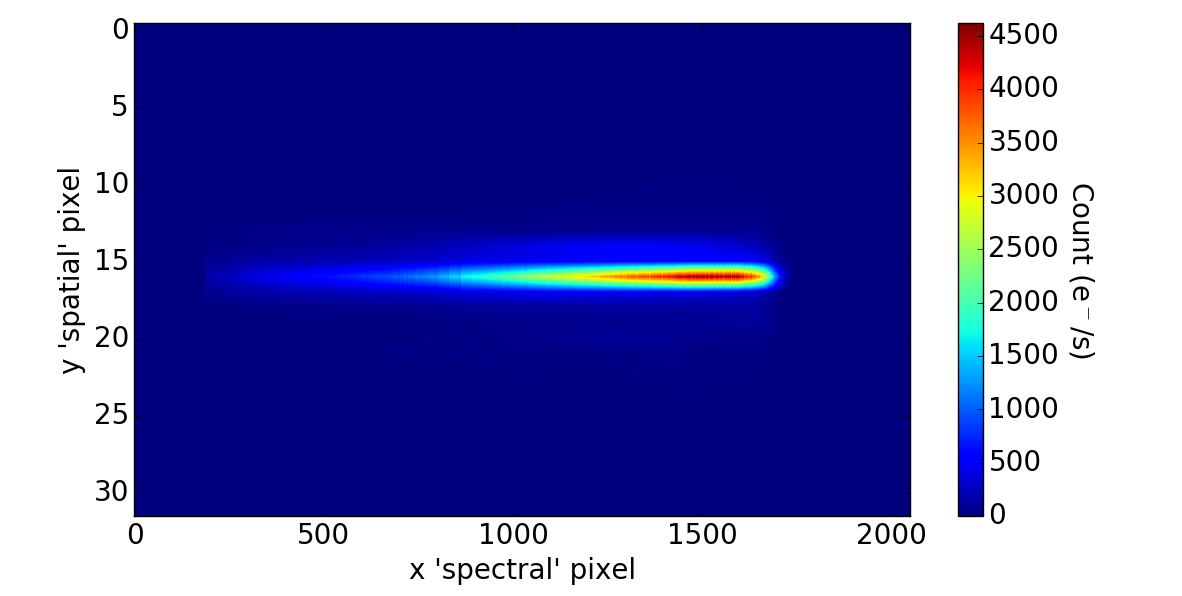}
 \end{center} 
    \caption{NIRSpec G395M/F290LP focal plane image simulated in JexoSim}
    \label{fig:NIRSPEC-fp}
\vspace{-0.2cm}    
\end{figure}

\begin{figure}
 \begin{center}
 	\includegraphics[width=1.0\columnwidth]{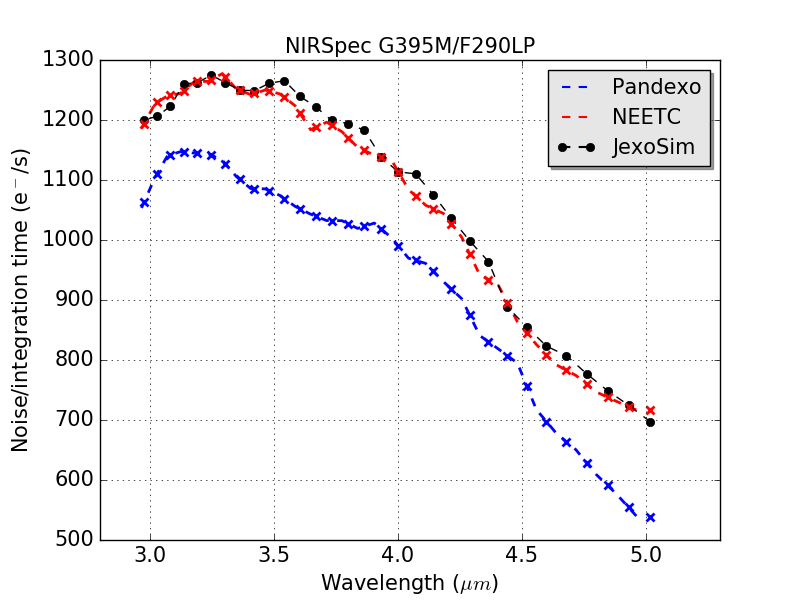}
 \end{center} 
    \caption{NIRSpec G395M/F290LP out-of-transit photon noise results obtained by JexoSim compared to Pandexo and NEETC results from B17. Binned $R$ power in JexoSim is 58.}
    \label{fig:NIRSPEC}
\end{figure}

\begin{figure}
 \begin{center}
 	\includegraphics[trim={1cm 0cm 2cm 0cm}, clip, width=1.0\columnwidth]{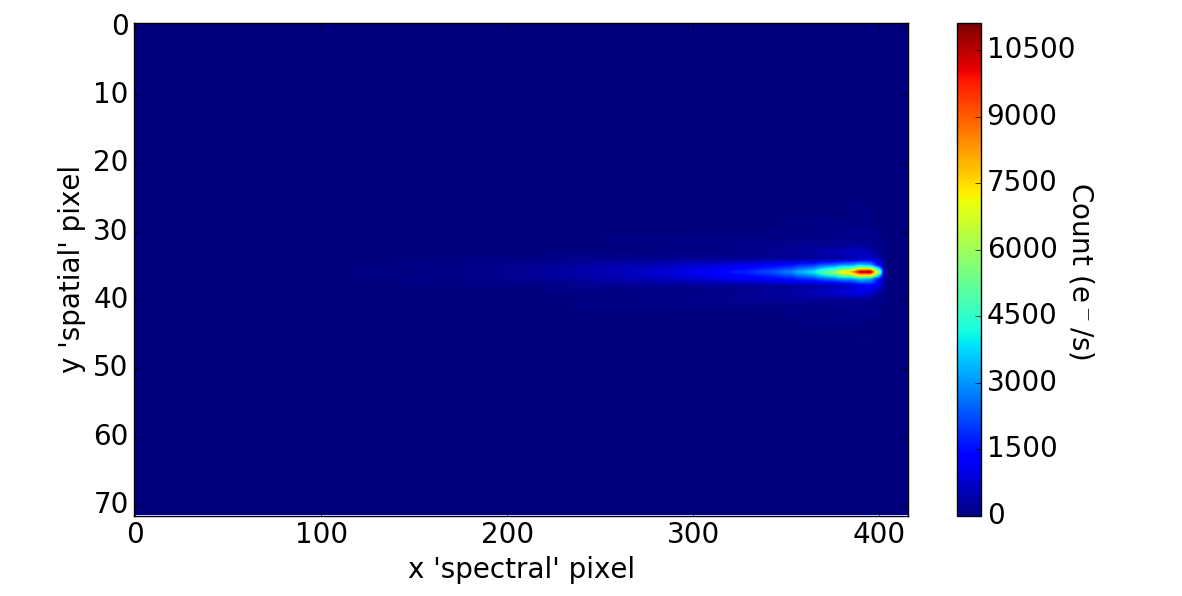}
 \end{center} 
    \caption{MIRI LRS focal plane image simulated in JexoSim}
    \label{fig:MIRI-fp}
\end{figure}

\begin{figure}
\vspace{1pt}
 \begin{center}
 	\includegraphics[width=1.0\columnwidth]{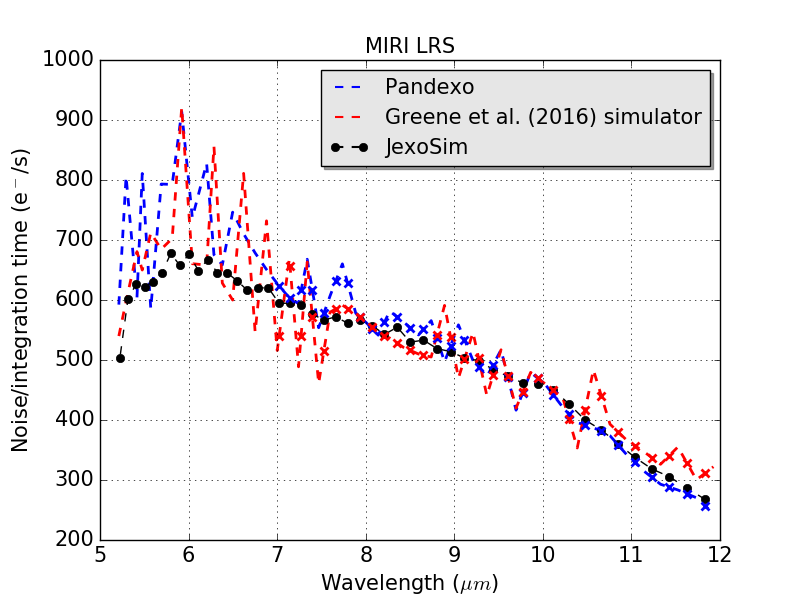}
 \end{center}  
    \caption{MIRI LRS out-of-transit photon noise results obtained by JexoSim compared to Pandexo and \protect\cite{Greene2016} simulator results from B17. Binned $R$ power in JexoSim is 58.}
    \label{fig:MIRI}
\end{figure} 

\subsection{NIRSpec}
The NIRSpec high throughput bright object time-series (BOTS) observing mode is optimized for transit spectroscopy of exoplanets.  This uses the S1600A1 1.6" x 1.6" aperture, which we model as a 16 pixel wide slit projection on the focal plane array ($W_s$ = 16).  For this paper, we simulate the G395M grism which has an intrinsic $R$ power of $\sim$ 1000 \citep{NIRSpec2019}, with the F290LP filter, which gives a wavelength range of 2.87-5.1 \textmu m. The SUB2048 subarray option is needed to record the full wavelength range.  Again matching the parameters used in B17, we use a LMF integration time, $t_{int}$, of 0.90156 s. The simulated focal plane from JexoSim is shown in Figure \ref{fig:NIRSPEC-fp}.  The spectra are binned to an $R$ power of 58 based on measurement of the distributions of points in B17. The noise results are shown in Figure \ref{fig:NIRSPEC}.  Notably there is much better agreement between the NEETC and JexoSim than between the NEETC and Pandexo.  When comparing points at the JexoSim spectral bin wavelengths, JexoSim has an average difference from the NEETC  of just +1.0 $\pm$ 1.5 \%, with a maximum deviation of +3.5 \%.  In comparison, Pandexo has an average difference from the NEETC of -12.6 $\pm$ 3.9 \% and a maximum deviation of -24.9 \%.  JexoSim deviates from Pandexo on average by +15.8 $\pm$ 5.6 \%, with a maximum difference of +30.7 \%.
It is not immediately apparent why there is a discrepancy with Pandexo compared to the NEETC result.  The agreement between JexoSim and the NEETC would tend to cross-validate these two simulators.

\subsection{MIRI}
The MIRI LRS mode uses a prism with a wavelength range of 5-12 \textmu m and an intrinsic $R$ power varying from 40 at 5\textmu m to 160 at 10 \textmu m \citep{MIRI2019}. Following B17, we use JexoSim to simulate the MIRI LRS slitless mode with the SLITLESSPRISM subarray, and use a $t_{int}$ of 1.431 s per LMF integration. The simulated focal plane image from JexoSim is shown Figure \ref{fig:MIRI-fp}. The 1-D spectra were binned to $R$ of 58, again based on our measurement of the distribution of points in B17. The noise results are shown in Figure \ref{fig:MIRI}. The Pandexo and \cite{Greene2016} MIRI simulator results at short wavelengths appear jagged which is attributed in B17 to binning methodology resulting in variable pixels per bin. The JexoSim Default Pipeline uses subpixelisation to produce smoothly varying $R$-bin sizes, and thus does not result in jagged binning at the shorter wavelengths.  We compare points at JexoSim spectral bin wavelengths above 8 \textmu m.  Over this wavelength range, JexoSim has an average difference of -1.0 $\pm$ 4.0 \% with Pandexo and -1.4 $\pm$ 6.3 \% with the \cite{Greene2016} MIRI simulator, and a maximum deviation of -10.6 \% from Pandexo and -14.1 \% from the \cite{Greene2016} simulator.  We consider this a good agreement with the previous simulators for wavelengths above 8 \textmu m.  JexoSim (through the JexoSim Default Pipeline) produces smoother $R$-binning than these other simulators which results in greater deviation for wavelengths below 8 \textmu m.

\section{Noise budget simulations}
\label{section: noise budget section}

In this section we demonstrate the capability of JexoSim for noise budget simulations.  In JexoSim, the noise contribution of different noise sources can be individually eludicated by running the simulation repeatedly with different noise sources switched on and off. This can reveal the relative contributions of different noise sources, providing a measure of performance, and may guide the optimal data reduction methods used. An out-of-transit simulation suffices for this kind of analysis. Since correlated noise integrates down differently to uncorrelated noise,  relative noise is best measured for the integrated signal at the time scale of the planet transit, $T14$.  To do this we use the Allan deviation method descibed in Section \ref{sec: Allan}. We simulate the fractional noise at $T14$ for two well-studied systems: GJ 1214 b, a super-Earth orbiting an M dwarf star, and HD 209458 b, a hot Jupiter orbiting a G type star.

\subsection{GJ 1214 b simulation}

To simulate observations for the super-Earth GJ 1214 b, we used a host stellar temperature of 3100 K, log $g=5$ and Fe/H = 0 to select the model PHOENIX  spectrum and an ecliptic latitude of 27.9$^\circ$ to obtain the $\beta$ coefficient for the zodi calculation.  The remaining system parameters were those in the Open Exoplanet Catalogue.  The transit duration, $T14$, calculated through JexoSim, was 3143 s.
We simulate NIRSpec and MIRI in the configurations given in Table \ref{table:Config}. Using the same random seed, we repeat simulations with different noise sources activated.  Each simulated out-of-transit observation is 3 hours long in simulated time. For NIRSpec we use $t_g=0.902$ s, and for MIRI we use $t_g=0.159$ s. The reset time ($t_{dead}$) and the duration of the zeroth read ($t_{zero}$), are both assumed to equal $t_g$. The number of groups, $n$, is determined using the minimum saturation time of a pixel on the array (assuming 100\% full well) (Eq. \ref{eq: n1}).  This gave $n$ of 61 with $t_{int}$ of 9.54 s in MIRI, and an observing efficiency of 96.8\%.  In NIRSpec, $n$ = 8, $t_{int}$ = 6.314 s, and the observing efficiency was 77.7\%.  To improve computational speed, we elected to perform last-minus-first processing, i.e. correlated double sampling (CDS).  This meant that only two images per integration needed to be generated (the zeroth and final reads) rather than the full stack of NDRs. This however will impact the read noise adversely relative to using up-the-ramp fitting or Fowler sampling.  If Fowler-8 sampling is performed in MIRI, the read noise would be reduced by a factor of $1/\sqrt{8}$ compared to the CDS noise obtained here.  Each 3 hour timeline is processed through the JexoSim Default Pipeline ending at step 8.  The $R$-binning chosen in each channel was the same as used in Section \ref{section: fp sims}. The noise sources simulated were as follows: photon noise (from the stellar target), dark current noise, zodiacal light noise, optical emissions noise, read noise, spatial jitter noise, spectral jitter noise, combined (both spectral and spatial) jitter noise and all noise sources combined.  The JDP steps used depended on the noise sources simulated.  If no jitter noise was simulated then jitter decorrelation was not applied. For the diffuse sources, dark current, emissions and zodi, the stellar signal was not simulated, and in data reduction the aperture mask was centred on the middle row of each image.
The processed out-of-transit light curves were then subjected to Allan deviation analysis to obtain the fractional noise in ppm at $T14$.  Parallel simulations using a noiseless stellar signal were used to obtain $\mu_{ \left\langle s \right\rangle }$ required for this analysis.  20 JexoSim realizations were performed for each noise source and an average value of the fractional noise obtained in each case.

\subsection{HD 209458 b simulation}
\label{noise:budget H209}

We also simulated the hot Jupiter HD 209458 b. We used a host stellar temperature of 6100 K, log $g=4.5$ and Fe/H = 0 for the model PHOENIX  spectrum, normalised to a J magnitude of 6.591.  An ecliptic latitude of 28.7$^\circ$ was used to obtain the $\beta$ coefficient for the zodi model.  For consistency with the spectrum simulation in Section \ref{section: HD 209458 b simulation}, the following parameters were used: $M_p=0.71$ $M_{J}$, $R_p= 1.31$ $R_{J}$ and $R_s= 1.2$ $R_{Sun}$. The remaining system parameters were taken from the Open Exoplanet Catalogue. $T14$ calculated through JexoSim was 10752 s. We set $t_g$, $n$ and $t_{int}$ 
for NIRSpec and MIRI to be the same as those used in Section \ref{section: HD 209458 b simulation}. As in Section \ref{section: HD 209458 b simulation}, for spectral binning, we used 5 pixel-wide bins for MIRI, and 30 pixel-wide bins for NIRSpec.  
The remaining aspects of the simulation and processing are the same as for GJ 1214 b above.

\begin{figure*}
\begin{center}
    \includegraphics[trim={0, 0.5cm, 0, 0}, clip,width=1.0\columnwidth]{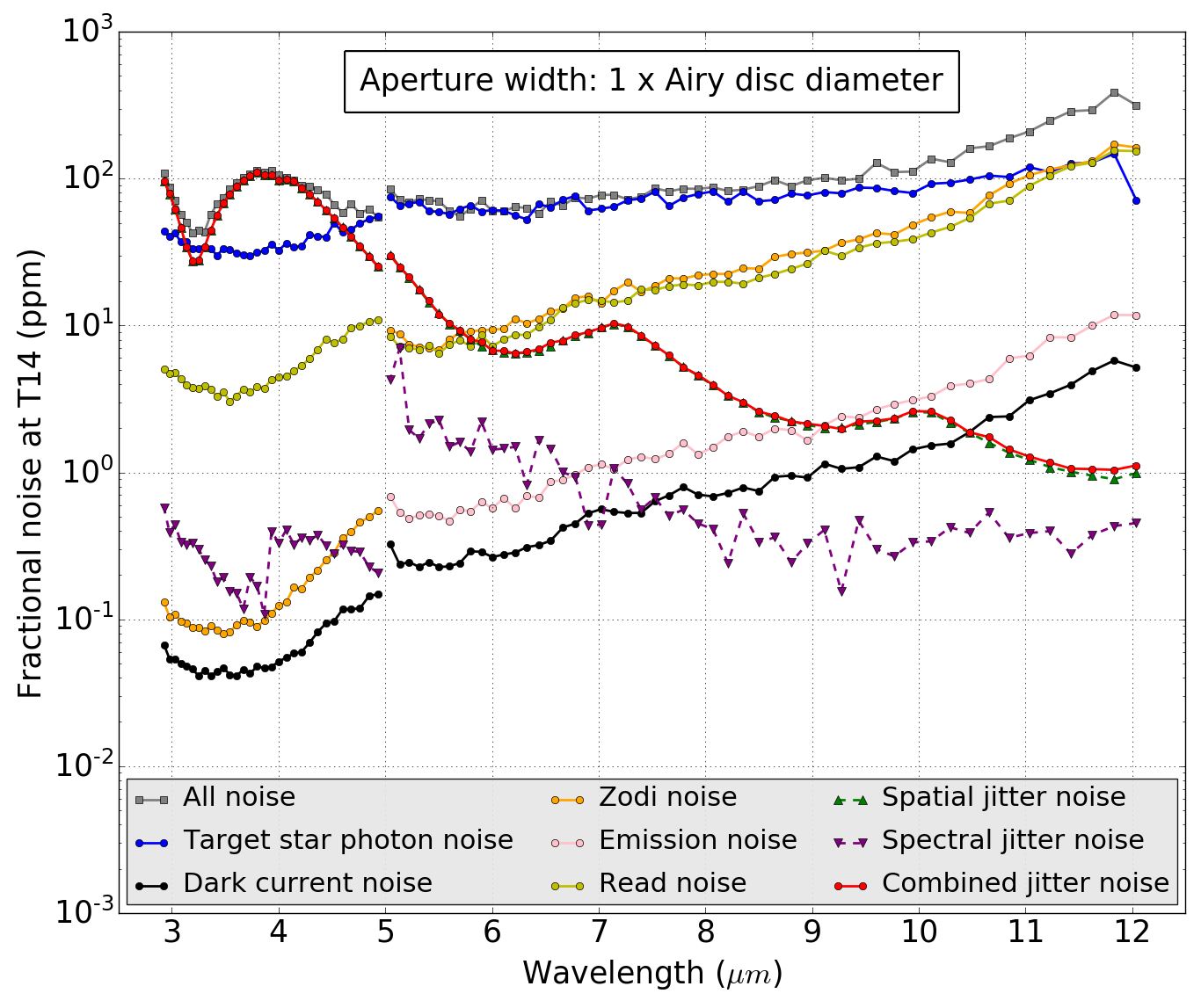}
 \includegraphics[trim={0, 0.5cm, 0, 0}, clip,width=1.0\columnwidth]{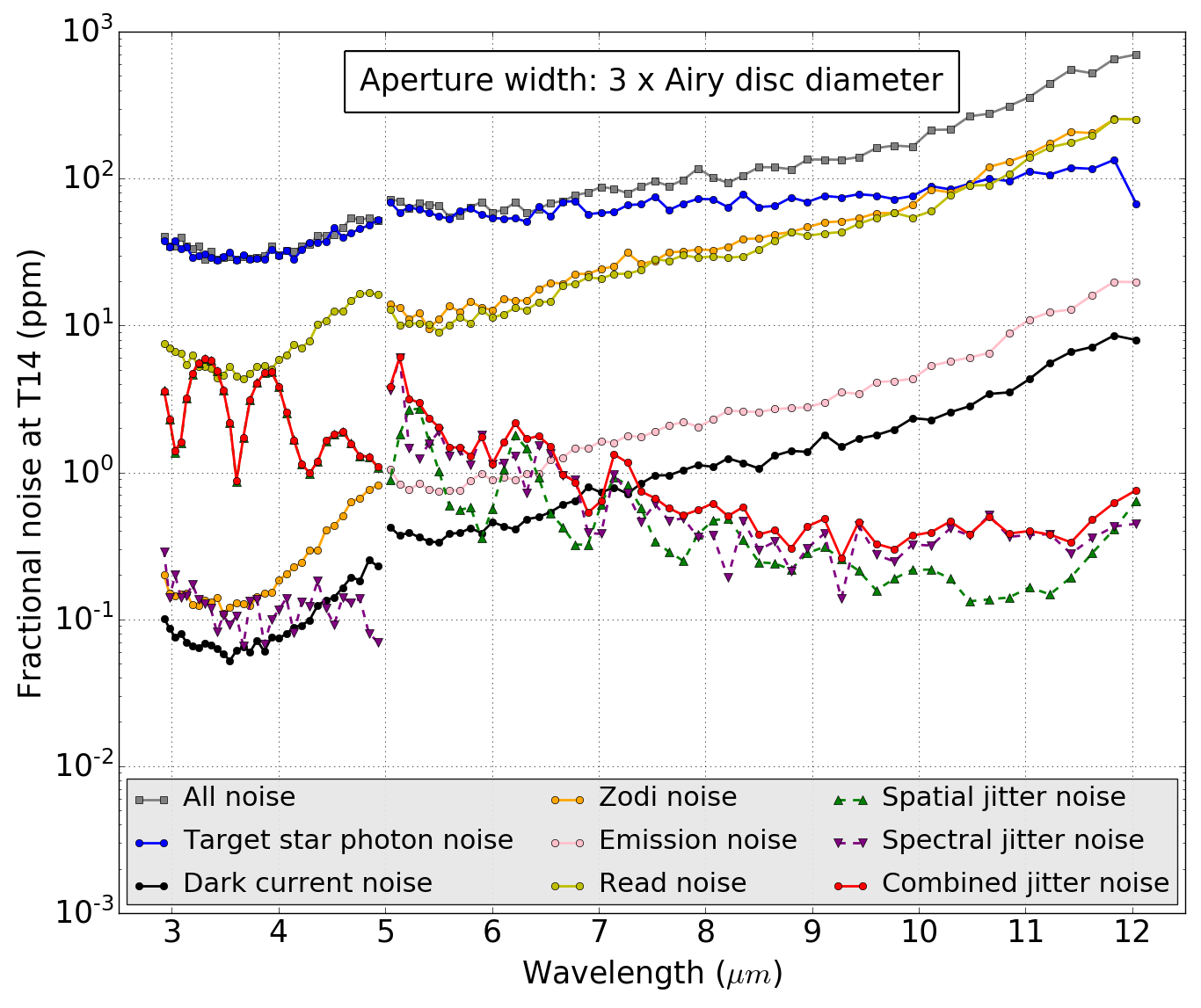}    
\end{center}    
{A: GJ 1214 b}
 
\begin{center}
    \includegraphics[trim={0, 0.5cm, 0, 0}, clip,width=1.0\columnwidth]{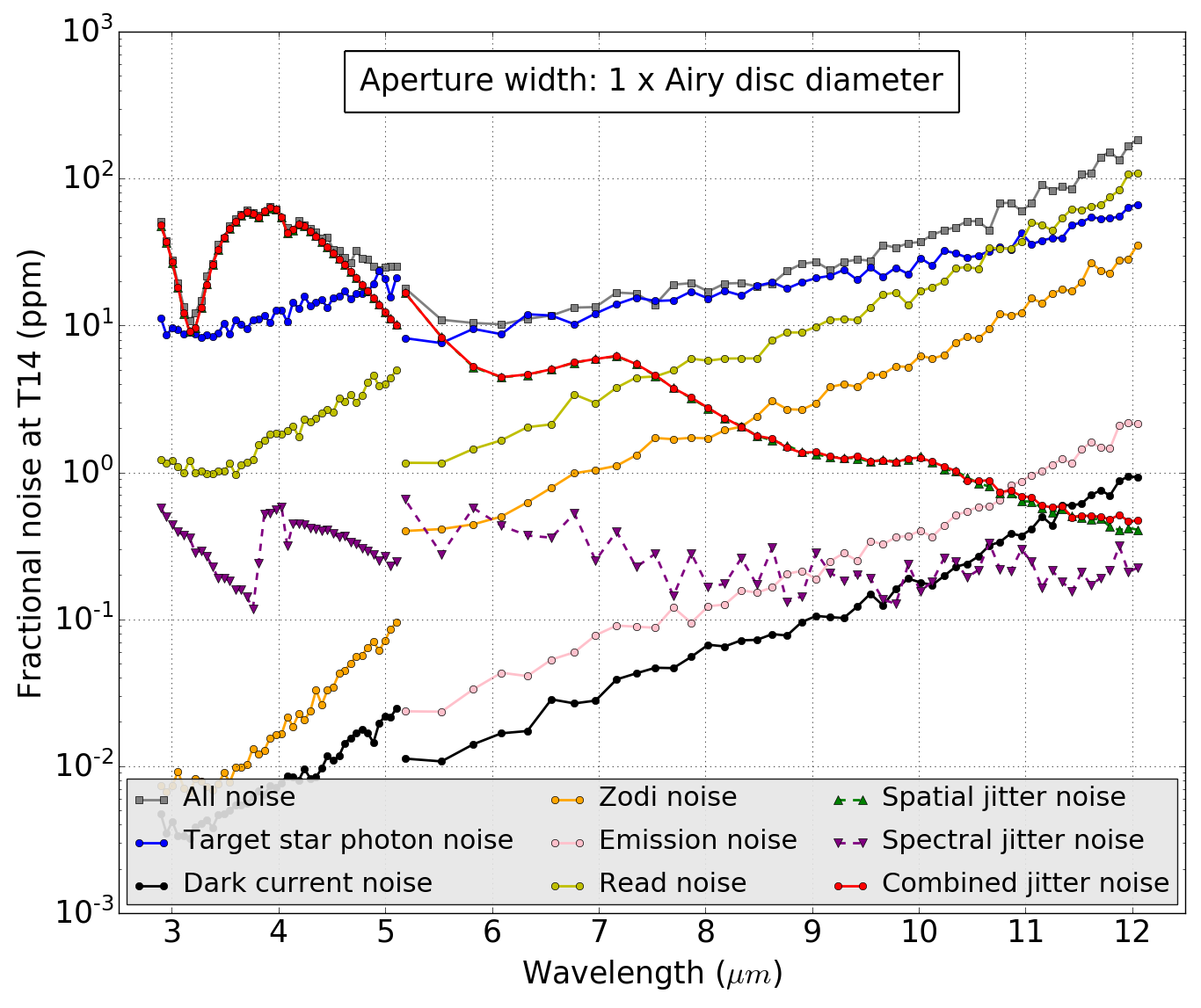}
 \includegraphics[trim={0, 0.5cm, 0, 0}, clip,width=1.0\columnwidth]{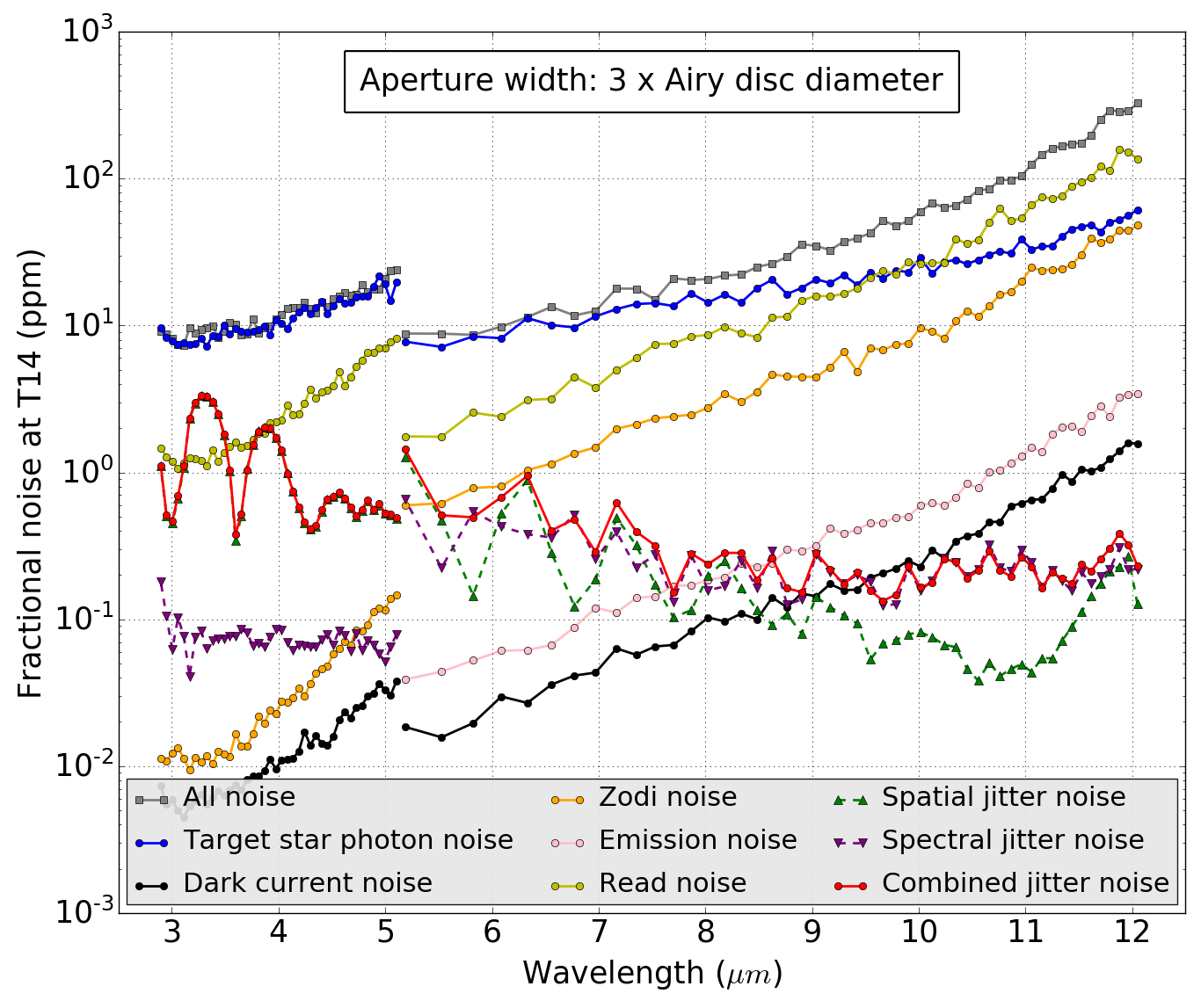}    
\end{center}      
{B: HD 209458 b}

\caption{Noise budgets obtained by JexoSim for a simulated observations of A) GJ 1214 b and B) HD 209458 b, using NIRSpec G395M/F209LP and MIRI LRS slitless mode, giving the fractional noise at the time scale of the transit duration, $T14$. Binning details are given in the text. Results shown are the average of 20 realizations. Left: Aperture width of 1 $\times$ Airy disc diameter. Right: Aperture width of 3 $\times$ Airy disc diameter.}
\label{noise_budget}  
\end{figure*}

\subsection{Noise budget results}
\label{section: noise budget results}

Two different widths (in the spatial direction) for the aperture mask were tested to see which one was optimal for each channel. These were: a) 1 $\times$ Airy disc diameter size (2.44 $F \lambda$), and b) 3 $\times$ the Airy disc diameter size (7.32 $F \lambda$). The mask aperture width therefore varies per pixel column.  We used $F$= 7.14 for MIRI \citep{Kendrew2015}, and $F$= 5.6 for NIRSpec \citep{DeMarchi2011}.  The results are shown in Figure \ref{noise_budget} for both planets studied.

For GJ 1214 b (Figure \ref{noise_budget} A),  using the 
1 $\times$ Airy disc diameter aperture we find that spatial jitter noise (and hence also combined jitter noise) becomes significant in NIRSpec exceeding the stellar photon noise over half of NIRSpec's wavelength range. This is most likely due to limits on the accuracy of centering  the mask on the peak of the signal, which impacts more at the blue end of the channel where both the mask aperture width and the PSF narrow.  Also, due to convolution with the intra-pixel response function, there is a widening of the PSF compared to the unconvolved PSF.  The impact of this convolution on the PSF shape will be higher for the narrower PSFs that occur at the shorter wavelengths than at the longer wavelengths, causing them to deviate relatively more from the ideal Airy disc width.  Finally, the PSFs generated using WebbPSF will contain optical abberations that will also tend to cause widening compared to the Airy function, which in turn will exacerbate the mismatch between mask aperture width and PSF width.  Use of advanced spectral extraction methods such as optimal extraction may obviate the need for an aperture mask and hence this noise effect.  However a simple solution would be to widen the mask aperture.  We find that when using the widened aperture (3 $\times$ Airy disc diameter), the spatial jitter noise falls well below the stellar photon noise over the wavelength range examined.  Widening the aperture however results in more noise from read noise, dark current and backgrounds as more pixels are included in the unmasked area.  The impact however of these on NIRSpec is minimal, with the total noise being stellar photon noise limited over the whole wavelength range examined.  For MIRI LRS, the lack of a slit results in a significant amount of zodical light contamination.  The noise from this impacts on the red end of the channel. Even when applying the narrower aperture, the zodi noise exceeds the stellar photon noise at wavelengths > 11.5 \textmu m. The read noise also exceeds the photon noise above this wavelength. The zodi and read noise contributions thus raise the total noise somewhat above the photon noise limit in the long wavelength half of the wavelength range examined. We noted above that the read noise can be improved by Fowler sampling which we have not simulated here.  With the wider aperture, the impact of the zodi and read noise sources becomes even more pronounced.  Observations with MIRI LRS in slitless mode may thus have improved signal-to-noise using narrower rather than wider apertures in data reduction.  Also, the slitted LRS mode (not simulated in this paper) will have reduced zodi contamination which could improve the signal-to-noise at the red end of the channel. 

For HD 209458 b (Figure \ref{noise_budget} B), using the 1 $\times$ Airy disc diameter aperture, we again see significant spatial jitter noise in NIRSpec arising from interaction of the signal with the mask. This is completely mitigated by widening the aperture to 3 $\times$ Airy disc diameter, where we again achieve a photon-noise limited performance. In MIRI, with the 1 $\times$ Airy disc diameter aperture, contributions from read noise and zodi noise again cause an increase the total noise above the photon noise limit over the red half of the spectrum.  There is also a slight impact on the total noise from spatial jitter noise at the extreme blue end of the channel.  On widening the aperture to  1 $\times$ Airy disc diameter, this jitter noise is again mitigated, but the impact from read noise and zodi noise goes up. 
 
Comparing the two aperture masks used, we conclude that the narrower aperture is the more optimal of the two for MIRI, since it minimizes the impact of zodi and read noise which are the most significant noise contributions together with the target star photon noise, although this is at the cost of a small added impact from spatial jitter noise at the blue end for HD 209458 b.  The wider aperture is optimal for NIRSpec since it mitigates the spatial jitter noise, and other noise sources do not reach the level of the target star photon noise.

\section{Planet spectrum simulations}

Finally, we show the ability of JexoSim to generate the key data product of a transit spectroscopy observation, a transmission spectrum with error bars.  The simulated spectrum with error bars captures the final experimental biases and uncertainties that then feed into spectral retrieval codes and the final science conclusions. In this paper we wish to show JexoSim's capability to generate such spectra from simulated transit observations.  Error bars capturing the precision can be obtained in a number of ways as discussed in Section \ref{sec 3.2}. Possibly the most robust and accurate way of determining the error bar is through multi-realization Monte Carlo simulations, which is the method we adopt in this section.  We simulate a primary transit spectroscopic observation of the hot Jupiter HD 209458 b using NIRSpec G395M/F290LP and MIRI LRS  in the configurations given in Table \ref{table:Config}.  We show the final spectrum with its error bars, and 
compare this to an equivalent spectrum from Pandexo v1.0 (using the Pandexo web interface\footnote{https://exoctk.stsci.edu/pandexo/}).
 
\begin{figure}
\begin{center}
    \includegraphics[trim={0 0 0 0}, clip,width=1.0\columnwidth]{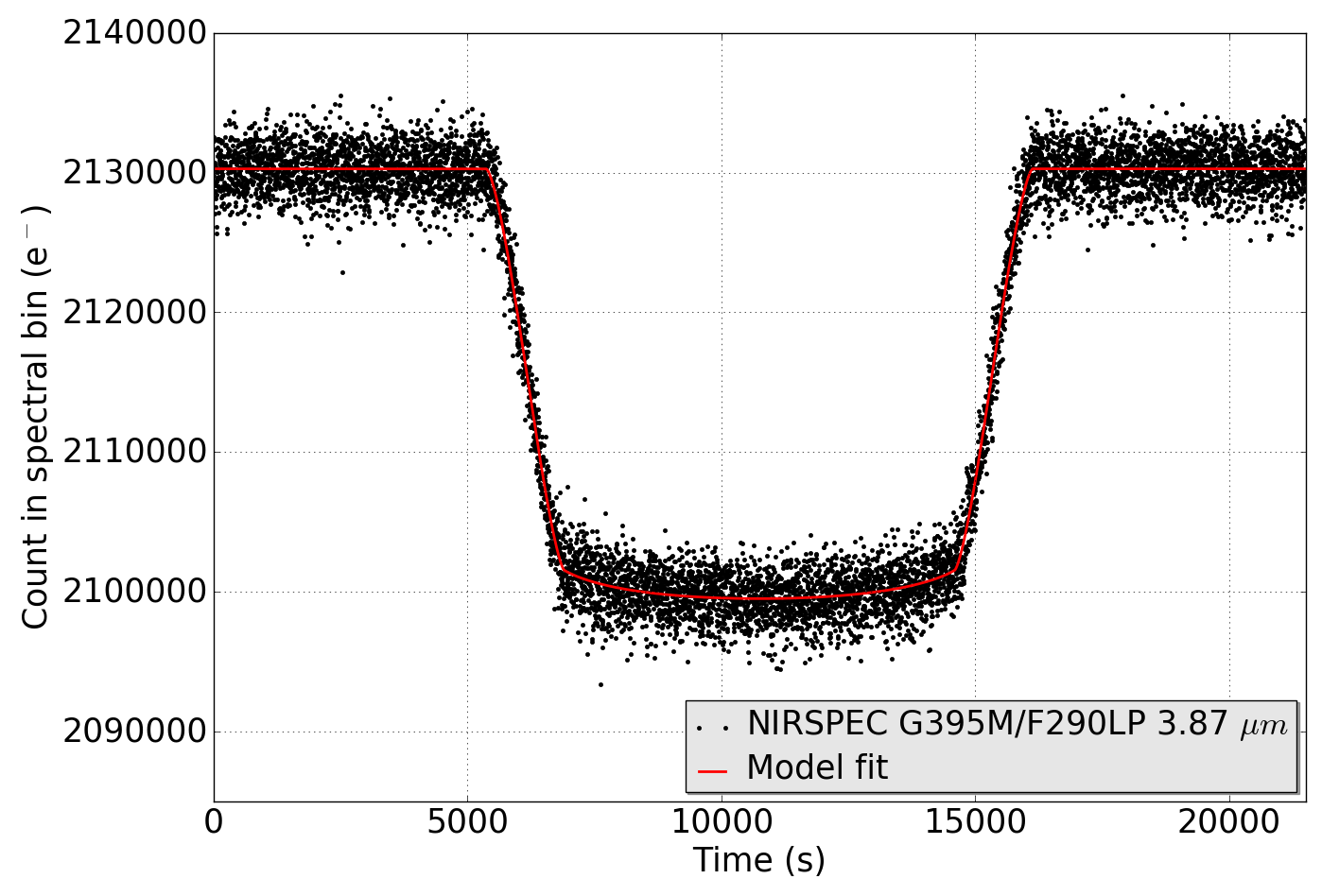}
{ A: NIRSpec}
    \end{center}

\begin{center}
 \includegraphics[trim={0 0 0 0}, clip,width=1.0\columnwidth]{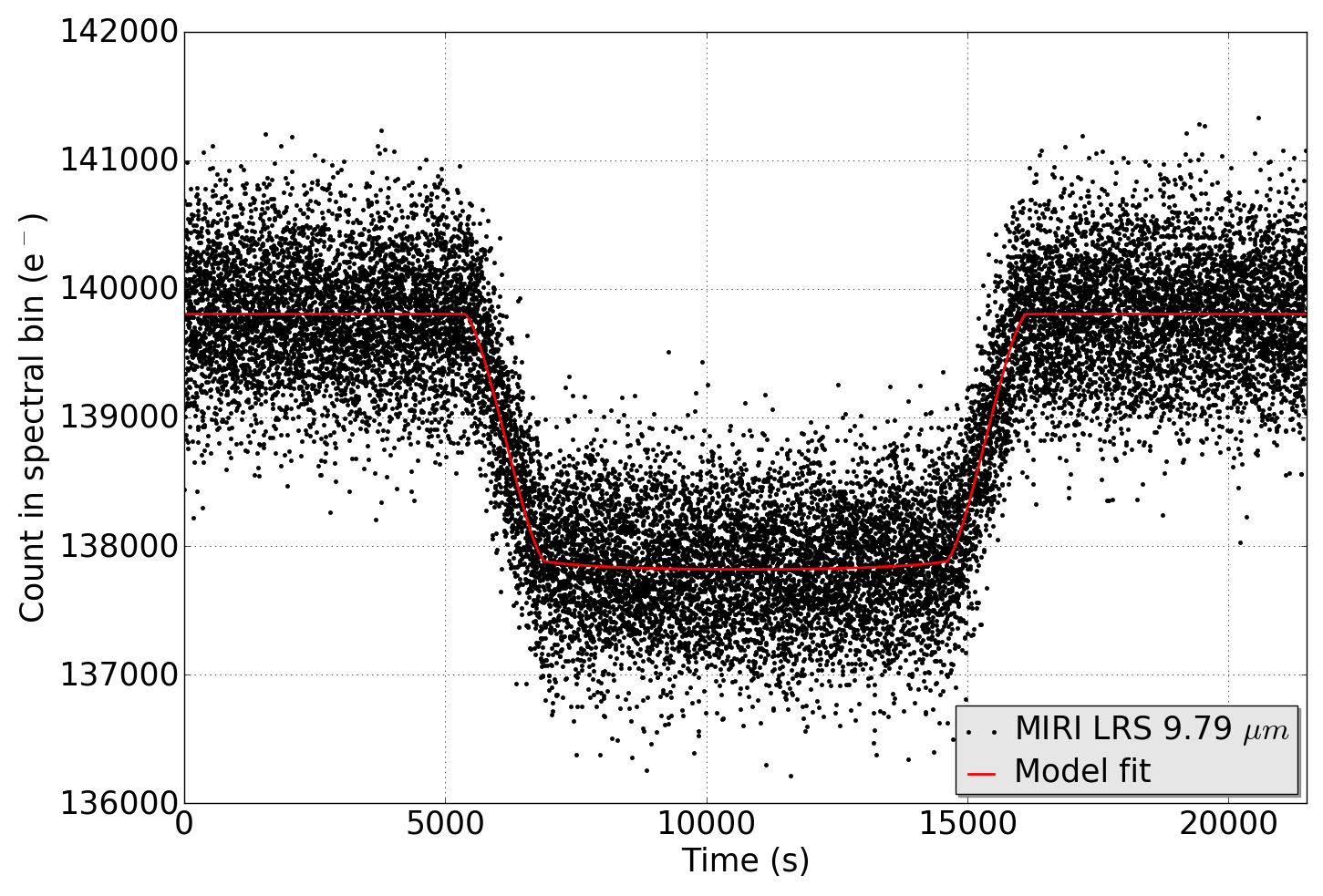}
{ B: MIRI}
 \end{center}
   
\caption{Examples of JexoSim spectral light curves obtained from a simulated transit spectroscopy observation of HD 209458 b. A: NIRSpec G395M/F290LP consisting of 7947 integrations. B: MIRI LRS slitless mode consisting of 19316 integrations. The red lines are the fitted Mandel-Agol models using quadratic limb darkening coefficients.}
\label{fig:light_curve}  
\end{figure}

\begin{figure*}
 \includegraphics[trim={0, 0cm, 0, 0cm}, clip,width=2.0\columnwidth]{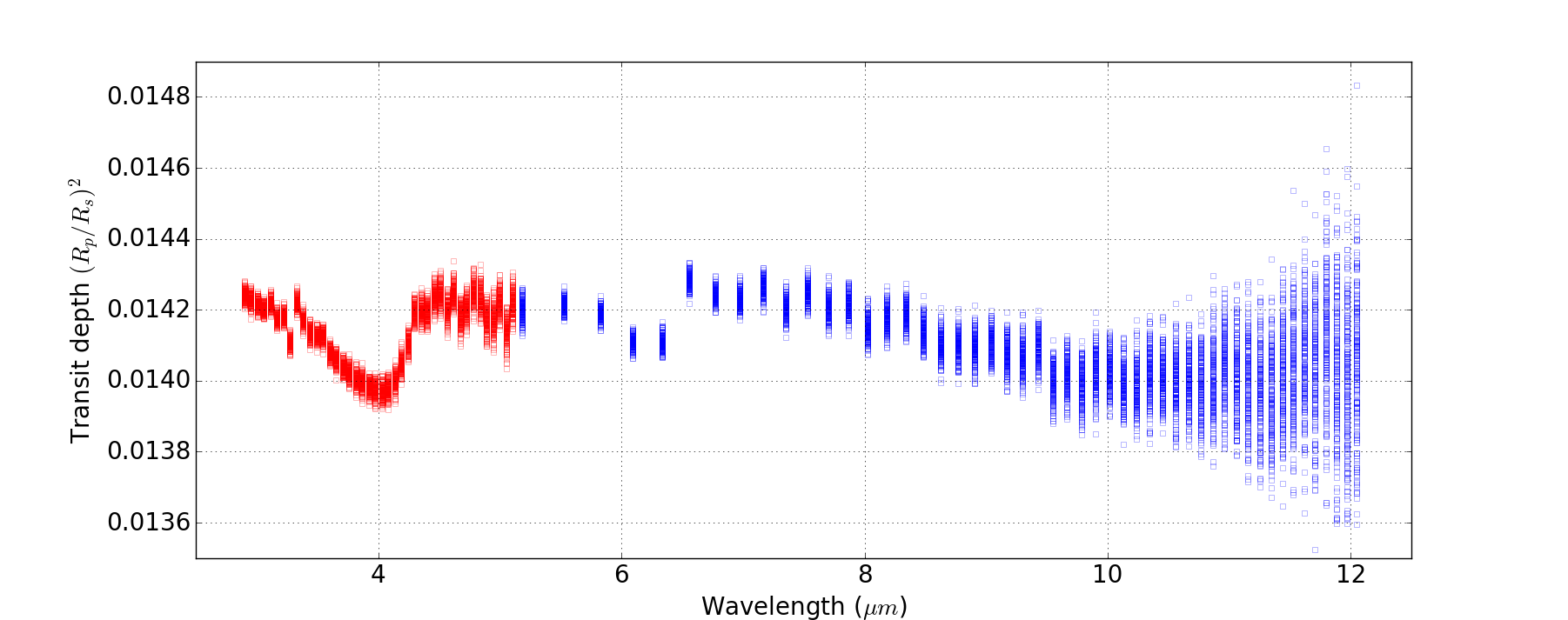}
\caption{Results of JexoSim Monte Carlo simulation of HD 209458 b transit observation. 200 realizations were performed for each spectral bin. Each data point shows the result of a single realization.  Red squares: NIRSpec. Blue squares: MIRI. The standard deviation of the distribution of transit depths for each spectral bin is used for the 1$\sigma$ error bar.}
\label{MC}  
\end{figure*}

\begin{figure*}
\centering
 \includegraphics[trim={0, 0cm, 0, 0cm}, clip,width=2.0\columnwidth]{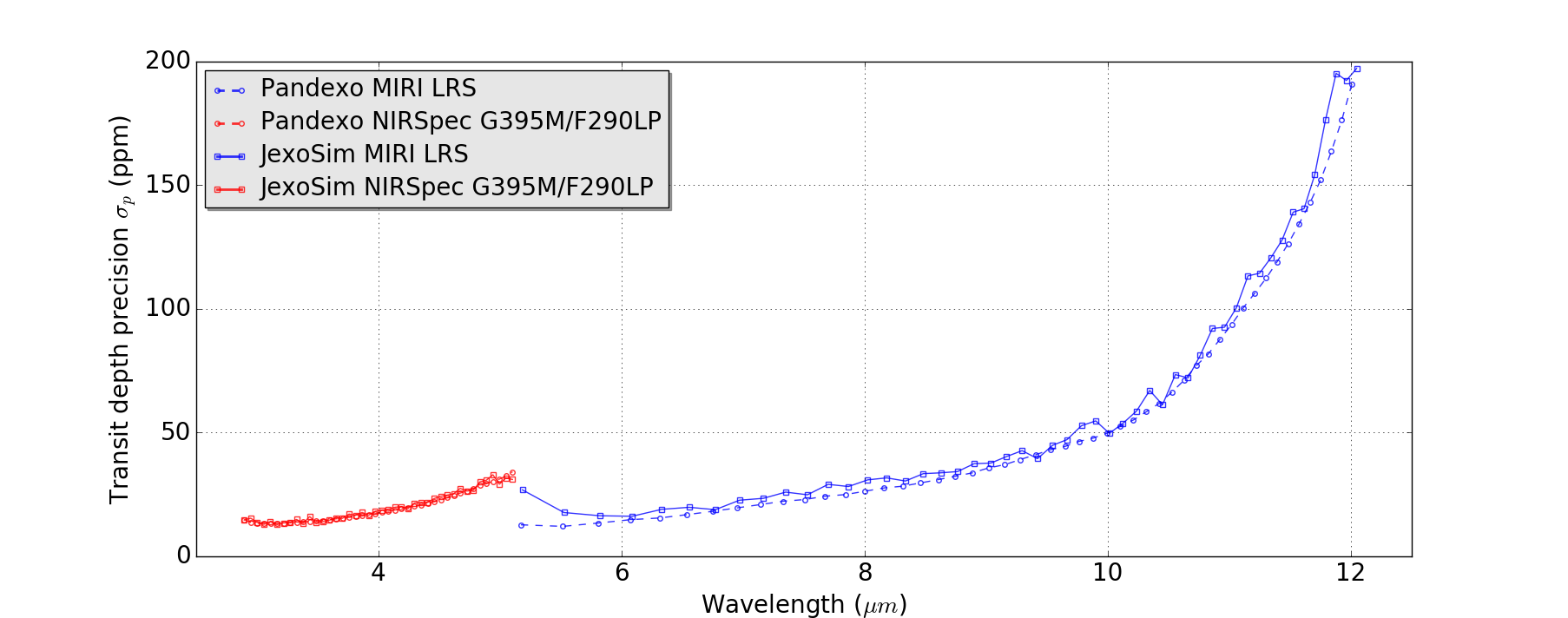}
\caption{Comparison of JexoSim and Pandexo transit depth uncertainty, $\sigma_p$, for simulated observation of HD 209458 b. For Pandexo, one data point at 4.67 \textmu m returned an erroneous value for the precision that was highly inconsistent with the values from the surrounding bins.  We therefore obtain the value for this point from an average of the values for the two adjacent points.}
\label{precision}  
\end{figure*}

\begin{figure*}
\begin{center}
 
    \includegraphics[trim={0cm, 0cm, 0, 1cm}, clip,width=2.0\columnwidth]{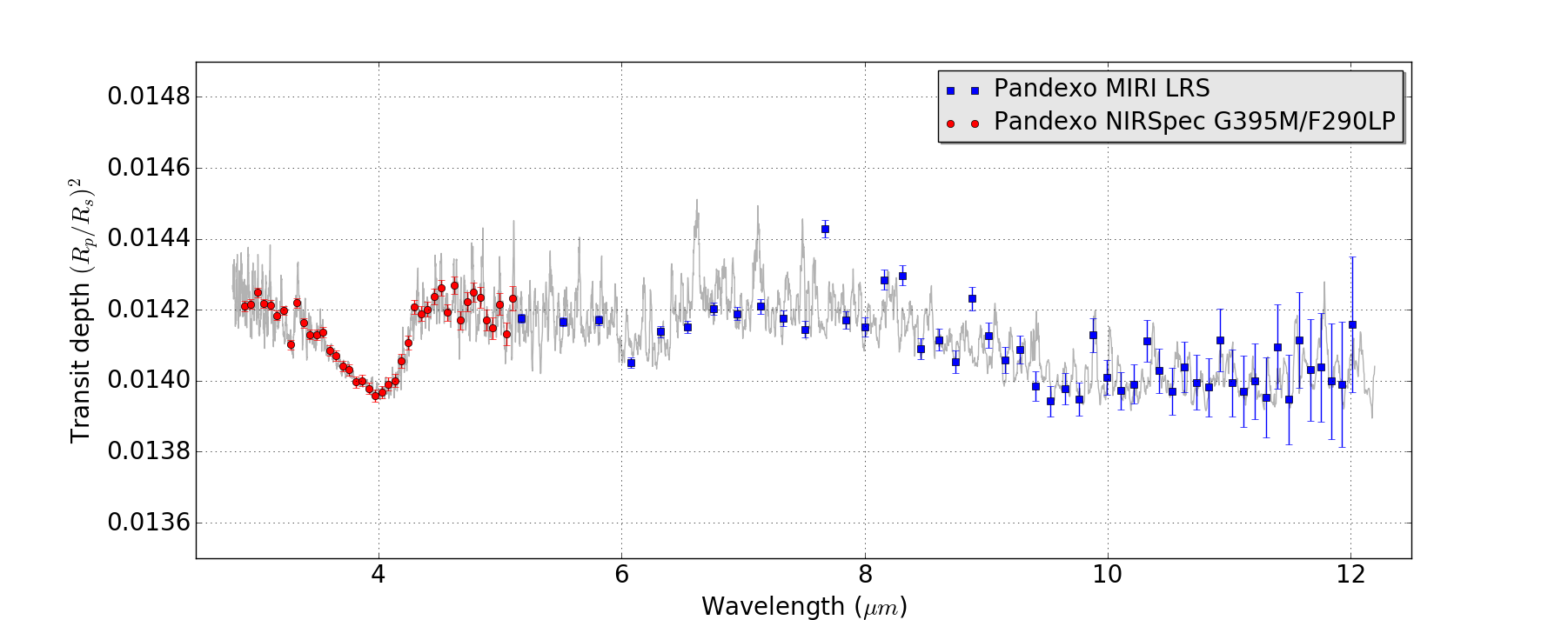}
    \end{center}
{A: Pandexo}

\begin{center}
 
 \includegraphics[trim={0cm, 0cm, 0, 1cm}, clip,width=2.0\columnwidth]{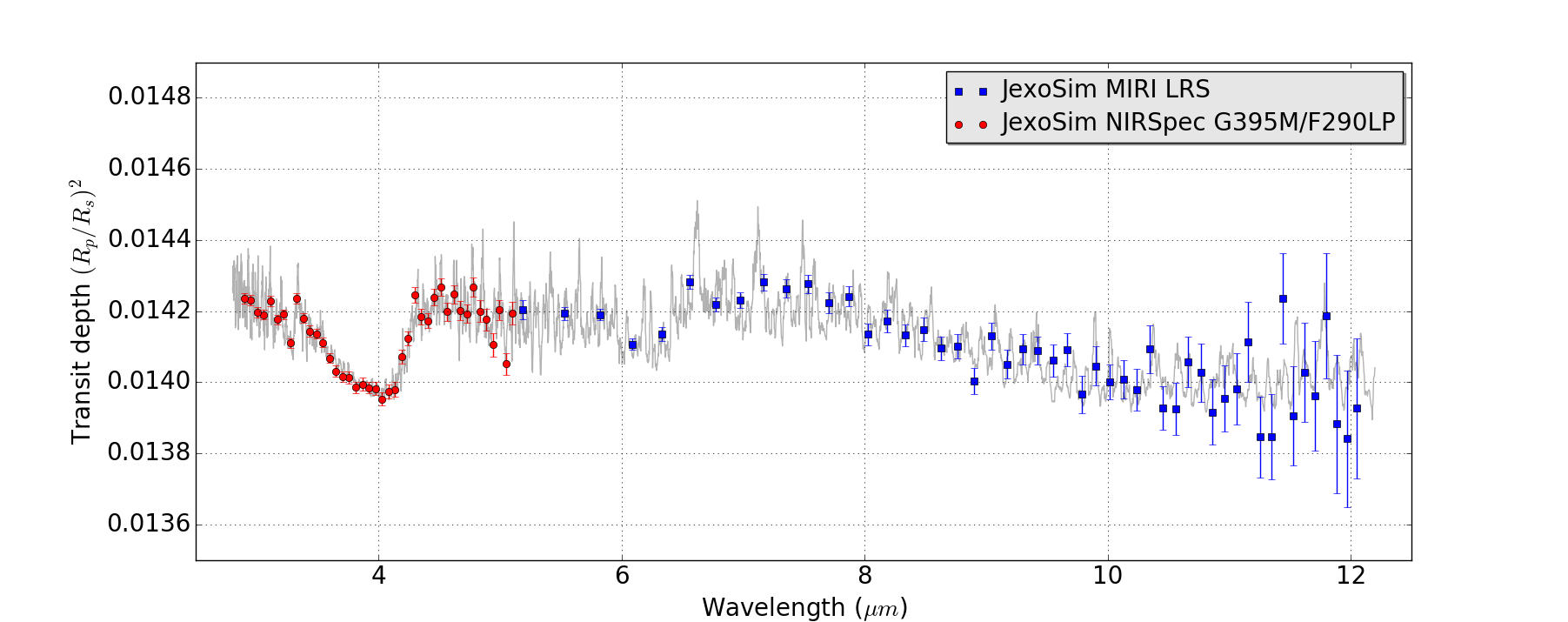}
 \end{center}
{B: JexoSim}

\caption{Simulated transmission spectrum for HD 209458 b: A) produced by Pandexo, B) produced by JexoSim. The grey line shows the input model spectrum (which has been smoothed for clarity). }
\label{spectrum}

\end{figure*}

\subsection{HD 209458 b simulation}
\label{section: HD 209458 b simulation}
For both Pandexo and JexoSim we utilised a model PHOENIX  spectrum with stellar temperture of 6100 K, log $g=4.5$ and Fe/H = 0, normalised to a J magnitude of 6.591. A model planet spectrum was generated using a Fortney model from the Exoplanet Characterization Toolkit\footnote{https://exoctk.stsci.edu/} and used for both simulators. This spectrum was generated for a planet temperature of 1000K, with equilibrium chemistry, and without TiO, clouds or scattering and the following parameters were used: $M_p=0.71$ $M_{J}$, $R_p= 1.31$ $R_{J}$ and $R_s= 1.2$ $R_{Sun}$.  These parameters were also adopted in the exosystem model in JexoSim, together with an ecliptic latitude of 28.7$^\circ$.  All remaining system parameters were taken from the Open Exoplanet Catalogue. $T14$ calculated through JexoSim was 10752 s, and this was also used as input for Pandexo.  In Pandexo, the `optimization' mode was used, with no addition noise floors added and 100\% saturation level chosen for detectors. For NIRSpec BOTS G395M/F290LP, Pandexo used $t_g$ = 0.902 s, returning $n$ = 2, giving a $t_{int}$ of 0.902 s, and duty cycle efficiency of 33.3 \%.  For MIRI LRS, Pandexo used $t_g$ = 0.15904 s, returning $n$ = 6, giving a $t_{int}$ of 0.7952 s, with a duty cycle efficiency of 71.4\%.  These same values for $t_g$,  $n$, and $t_{int}$ were then used in JexoSim.  JexoSim simulated a total observing time equivalent to 2 $\times$ $T14$, producing 7947 integrations for NIRSpec and 19316 integrations for MIRI.
According to B17, Pandexo simulates only photon noise from the stellar target, read noise and zodi noise, and performs last-minus-first processing.   
In JexoSim, we performed 200 realizations of the transit observation starting with the same random seed for each channel. All noise sources and backgrounds were activated. Due to the very short integration times `instantaneous' light curves rather than `integrated' ones were used. The JexoSim Default Pipeline was used to process the time series images, using the complete algorithm from Section \ref{sec 3.2}.  To match the Pandexo output which gives options for pixel column binning but not $R$-binning, fixed bins of a given number of pixel columns were used.  We used 5 pixel-wide bins for MIRI, and 30 pixel-wide bins for NIRSpec.  Based on the results in Section \ref{section: noise budget results}, we used mask apertures of width equal to 1 $\times$  Airy disc diameter for MIRI and 3 $\times$ Airy disc diameter for NIRSpec.  
The MIRI aperture is thus kept to a minimum to limit the effects of read noise and zodi noise at the red end of the channel.  
For NIRSpec, these noise sources are not significant according to the noise budget study, allowing for a wider aperture, which in turn is needed to reduce the spatial jitter noise.  

For this Monte Carlo simulation we used a modified version of the code that increases computational speed. This involves cropping the focal plane array to a smaller size that contains the entire spectrum, but has reduced background regions, before applying noise. Background counts are obtained using a separate array representing only the edges of the focal plane, and assume no significant stellar signal in these regions. The mask aperture width remains less than the width of the cropped focal plane, and the accuracy of results should not be affected by this modification.

Matching Pandexo, the JDP performs only last-minus-first processing (i.e. correlated double sampling), and thus overestimates the read noise compared to Fowler sampling or up-the-ramp slope fitting.  Model light curves were fitted to each binned spectral light curve to find $p(\lambda)$.  On fitting the model light curves only $p(\lambda)$ was a free parameter since we were interested in estimating just its uncertainty. The z-grid and limb darkening coefficients were thus kept fixed to those values used in the simulation.  Examples of spectral light curves and fitted models are shown in Figure \ref{fig:light_curve}.

\subsection{Transmission spectrum}

From the Monte Carlo simulation we obtain 200 transit depths per spectral bin, which are shown in Figure \ref{MC}. The standard deviation of the distribution in each bin gives the 1$\sigma$ uncertainty for a single transit depth measurement.  In Figure \ref{precision}, the 1$\sigma$ uncertainty, $\sigma_p(\lambda)$, which gives the size of the error bar on the transit depth, is shown for both JexoSim and Pandexo, both in good agreement.  Figure \ref{spectrum} B shows a JexoSim reconstructed transmission spectrum for HD 2098458 b.  The transit depths shown are for a single realization chosen at random from the 200 performed, and the error bars show the 1$\sigma$ uncertainty for each spectral bin.  The equivalent spectrum from Pandexo is shown in Figure \ref{spectrum} A.   Comparing the precision results in Figure \ref{precision}, we see a good agreement between JexoSim and Pandexo for NIRSpec.  $\sigma_p(\lambda)$ ranges from 12.8-32.9 ppm for JexoSim and 13.1-34.1 ppm for Pandexo.
This may be expected since photon noise from the stellar target will dominate in this channel.  However the error bars were obtained by two completely different methods: an estimate using the out-of-transit noise in Pandexo, and the Monte Carlo method using full transit simulations in JexoSim.   This validates the method used in JexoSim, but the Monte Carlo method has potentially more versatility for capturing the errors from complex noise sources and revealing systematic biases.  Comparing the precision obtained in MIRI, we again obtain a fairly good agreement across most of the wavelength band. $\sigma_p(\lambda)$ ranges from 16.1-197.2 ppm for JexoSim and 12.0-190.9 ppm for Pandexo. At the extreme red end of the wavelength range, the noise in JexoSim is slightly higher than in Pandexo.  This could be due to the different methods used to obtain the error bar in the two simulators, or possibly due to differences in the in the zodiacal light models used, since from the results of Section \ref{noise_budget} we know that both zodiacal light and read noise become significant noise sources in this wavelength region. Of note, the background noise model has been updated in PandExo/Pandeia v1.3, which increases noise at the red end, and which could make these precision estimates more compatible.
At the extreme blue end, the noise is slightly higher in JexoSim compared to Pandexo.  From the results in Section \ref{section: noise budget section}, this is likely to be due to spatial jitter noise as the edges of the aperture mask interact with the signal; this might respond to selective widening of the aperture at just the blue end of the channel or may disappear using maskless methods such as optimal extraction.
Although binned to the same number of pixel columns, there is a very small mismatch between the positions of the transit depth points in wavelength space between Pandexo and JexoSim for MIRI.  This might be due to differences in the starting pixel column chosen, or possibly the way the central wavelength is calculated for each bin.

\section{Conclusions}
We report JexoSim, a new simulation package for transit spectroscopy with JWST. JexoSim can model both instrumental and astrophysical processes, capturing complex noise and systematics in the time domain, permitting synthetic observations of transiting exoplanets. We validated JexoSim against Pandexo and independent instrument simulators, with good levels of agreement for photon noise.  We demonstrated how JexoSim can be used to estimate noise budgets revealing the noise contributions of different sources, which can be used for optimisation of observing and data reduction strategies. We find that zodiacal light and read noise are both significant noise sources in MIRI at longer wavelengths. We also find that if apertures are used, optimal sizing will be important to balance spatial jitter noise against read noise and background noise contributions. 

We simulated a transmission spectrum of the hot Jupiter HD 209458 b, generating time series spectral images akin to a real observation.  These were processed through a pipeline and model light curves fitted to reconstruct a spectrum. Using a Monte Carlo approach, we directly measured the probability distribution of the transit depths, thereby obtaining a robust estimate of the 1$\sigma$ error bars. Our method was validated by comparison with the precision estimated from Pandexo. The results given in this paper are valid only for the simulated observing conditions used, and we did not attempt to find the optimal observing strategy for any of the simulations.  Our results have also been dependent on use of the JexoSim Default Pipeline and the use of last-minus-first processing. Therefore the noise results shown here may be improved when optimal observing strategies are used, and when more advanced pipeline methods are considered. 

This baseline version of JexoSim will continue to be augmented and developed with further systematic processes added as more and up-to-date instrumental information is obtained.  This will further improve the accuracy of the simulation in JexoSim.  The accuracy of the JexoSim pointing simulation would also benefit from a PSD model derived from spacecraft modeling studies rather than the scaled version of the Herschel PSD used here.  Further optimisation of the pipeline will be performed. We envisage the data output from JexoSim will eventually feed directly into planned official pipelines for JWST instruments.  Overall, we have shown JexoSim to be a unique and versatile tool that will assist JWST in achieving its goal of transformational exoplanet science. 
 
\section{Acknowledgements}

We acknowledge the following software packages and databases required for this article. JexoSim utilized NumPy v1.10.2 \citep{Oliphant2006} , SciPy v0.15.1 \citep{Jones2001}, Matplotlib v2.2.4 \citep{Hunter2007}, pandas v0.16.2 \citep{McKinney2010}, ExoData v2.1.7 \citep{Varley2016}, PyTransit v1.0 \citep{Hannu2015}(doi:10.5281/zenodo.157363), Astropy v2.0.12 \citep{astropy2013,astropy2018}, PyFITS v3.5 (https://pypi.org/project/pyfits/3.5/) (PyFITS is a product of the Space Telescope Science Institute, which is operated by AURA for NASA), ExoTETHyS \citep{Morello2019}, WebbPSF v0.8.0 \citep{Perrin2012}, Pandeia database v1.3 \citep{Pontoppidan2016}, the Open Exoplanet Catalogue \citep{Rein2012}, and the PHOENIX BT-Settl database \citep{Allard2012}. The paper also utilized Pandexo v1.0 \citep{Batalha2017} (via the online interface), the Exoplanet Characterization Tookit (https://exoctk.stsci.edu/), and
Webplotdigitizer \citep{Webplot}.

S.S. and A.P. were supported by United Kingdom Space Agency (UKSA) grant: ST/S002456/1. We thank G. Morrelo (CEA-Saclay) for providing the ExoTETHyS code for calculation of limb darkening coefficients. We thank E. Pascale (La Sapienza University of Rome)  and M. Griffin (Cardiff University) for their advice and very helpful comments on this paper.  We thank the referee Natasha Batalha for her excellent and helpful comments.



\bibliographystyle{mnras}


\vspace{-0.5em}

\begin{thebibliography}{}
\makeatletter
\relax
\def\mn@urlcharsother{\let\do\@makeother \do\$\do\&\do\#\do\^\do\_\do\%\do\~}
\def\mn@doi{\begingroup\mn@urlcharsother \@ifnextchar [ {\mn@doi@}
  {\mn@doi@[]}}
\def\mn@doi@[#1]#2{\def\@tempa{#1}\ifx\@tempa\@empty \href
  {http://dx.doi.org/#2} {doi:#2}\else \href {http://dx.doi.org/#2} {#1}\fi
  \endgroup}
\def\mn@eprint#1#2{\mn@eprint@#1:#2::\@nil}
\def\mn@eprint@arXiv#1{\href {http://arxiv.org/abs/#1} {{\tt arXiv:#1}}}
\def\mn@eprint@dblp#1{\href {http://dblp.uni-trier.de/rec/bibtex/#1.xml}
  {dblp:#1}}
\def\mn@eprint@#1:#2:#3:#4\@nil{\def\@tempa {#1}\def\@tempb {#2}\def\@tempc
  {#3}\ifx \@tempc \@empty \let \@tempc \@tempb \let \@tempb \@tempa \fi \ifx
  \@tempb \@empty \def\@tempb {arXiv}\fi \@ifundefined
  {mn@eprint@\@tempb}{\@tempb:\@tempc}{\expandafter \expandafter \csname
  mn@eprint@\@tempb\endcsname \expandafter{\@tempc}}}



\bibitem[\protect\citeauthoryear{Allard, Homeier  \& Freytag}{Allard
  et al.}{2012}]{Allard2012}
Allard F.,  Homeier D.,   Freytag B.,  2012, \mn@doi [Philos. Trans. Royal Soc. A] {10.1098/rsta.2011.0269}, 370, 2765

\bibitem[\protect\citeauthoryear{{Astropy Collaboration}}{{Astropy Collaboration} }{2013}]{astropy2013}
Astropy Collaboration et al.,  2013, \mn@doi[A\&{A}]
  {10.1051/0004-6361/201322068}, 558, A33  

\bibitem[\protect\citeauthoryear{{Astropy Collaboration}}{{Astropy Collaboration} }{2018}]{astropy2018}
Astropy Collaboration et al.,  2018, \mn@doi[AJ]
  {10.3847/1538-3881/aabc4f}, 156, 123 
  

\bibitem[\protect\citeauthoryear{Barron et~al.,}{Barron
  et al.}{2007}]{Barron2007}
Barron N. et al., 2007, \mn@doi [PASP] {10.1086/517620}, 119, 466
 


\bibitem[\protect\citeauthoryear{Batalha, N. et al.}{Batalha, N. et al.}{2015}]{Batalha2015}
Batalha N.  et al., 2015, {preprint (\mn@eprint {} {arXiv:1211.7121})}


\bibitem[\protect\citeauthoryear{Batalha, N.E. et al.}{Batalha, N.E.
  et al.}{2017}]{Batalha2017}
Batalha N. E. et al., 2017, \mn@doi [PASP] {10.1088/1538-3873/aa65b0}, 129, 064501


\bibitem[\protect\citeauthoryear{Bean et~al.,}{Bean
  et al.}{2018}]{Bean2018}
Bean J. L. et al., 2018, \mn@doi [PASP] {10.1088/1538-3873/aadbf3}, 130, 114402


\bibitem[\protect\citeauthoryear{Beichman et al.}{Beichman et al.}{2012}]{Beichman2012}
Beichman C. et al., 2012, \mn@doi[Proc. SPIE] {10.1117/12.925447}, 84422N 

\bibitem[\protect\citeauthoryear{Beichman et al.}{Beichman
  et al.}{2014}]{Beichman2014}
Beichman C.  et al., 2014, \mn@doi [PASP] {10.1086/679566}, 126, 1134

\bibitem[\protect\citeauthoryear{Berta et al.}{Berta et al.}{2012}]{Berta2012}
Berta Z. K.  et al., 2012, \mn@doi [ApJ]
  {10.1088/0004-637x/747/1/35}, 747, 35


\bibitem[\protect\citeauthoryear{{Charbonneau}, {Brown}, {Noyes}  \&
  {Gilliland}}{{Charbonneau} et al.}{2002}]{Charb2002}
{Charbonneau} D. et al.,  2002,
  \mn@doi [ApJ] {10.1086/338770}, 568, 377
  
\bibitem[\protect\citeauthoryear{De Marchi}{De Marchi}{2011}]{DeMarchi2011}
De Marchi G. et al., 2011, \mn@doi[Proc. SPIE] {10.1117/12.893914}, 81500C
  
\bibitem[\protect\citeauthoryear{Deming et al.}{Deming et al.}{2013}]{Deming2013}
Deming D.  et al., 2013, \mn@doi [ApJ]
  {10.1088/0004-637X/774/2/95}, 774, 95 
  

 
\bibitem[\protect\citeauthoryear{Doyon et al.,}{Doyon et al.}{2012}]{Doyon2012}
Doyon R. et al., 2012, \mn@doi[Proc. SPIE] {10.1117/12.926578}, 84422R

\bibitem[\protect\citeauthoryear{Ferruit et al.,}{Ferruit et al.}{2014}]{Ferruit2014}
Ferruit, P. et al., 2014, \mn@doi[Proc. SPIE] {10.1117/12.2054756}, 91430A


\bibitem[\protect\citeauthoryear{Gardner et al.,}{Gardner
  et al.}{2006}]{Gardner2006}
Gardner J. P.  et al., 2006, \mn@doi [Space Science Reviews]
  {10.1007/s11214-006-8315-7}, 123, 485

\bibitem[\protect\citeauthoryear{Giardino et al.}{Giardino et al.}{2012}]{Giardino2012}
Giardino, G. et al., 2012, \mn@doi[Proc. SPIE] {10.1117/12.925357},  84531T 


\bibitem[\protect\citeauthoryear{Goudfrooij, Albert  \& Doyon}{Goudfrooij
  et al.}{2015}]{Goudfrooij2015}
Goudfrooij P.,  Albert L.,   Doyon R.,  2015, STScI Newsletter, 32(1), 37

\bibitem[\protect\citeauthoryear{Greene et al.}{Greene et al.}{2016}]{Greene2016}
Greene, T. P. et al., 2016, \mn@doi [ApJ]{10.3847/0004-637X/817/1/17}, 817, 17
 

\bibitem[\protect\citeauthoryear{Hardy, Willot \& Pazder}{Hardy, Willot \& Pazder}{2014}]{Hardy2014}
Hardy, T., Willot, C., Pazder, J.,  2014, \mn@doi[Proc. SPIE]{10.1117/12.2057114}, 91542D 



\bibitem[\protect\citeauthoryear{Hunter}{Hunter}{2007}]{Hunter2007}
Hunter J. D., 2007, \mn@doi [Comput. Sci. Eng.] {10.1109/MCSE.2007.55}, 9, 90


\bibitem[\protect\citeauthoryear{James et al.}{James et al.}{1997}]{James1997}
James J. F. et al.,  1997,
  \mn@doi [MNRAS]
  {10.1093/mnras/288.4.1022}, 288, 1022
  
\bibitem[\protect\citeauthoryear{Jones et al.}{Jones et al.}{2001}]{Jones2001}
Jones E. et al.,  2001, SciPy: Open Source Scientific Tools for Python {http://www.scipy.org/}
  

\bibitem[\protect\citeauthoryear{Kendrew et al.,}{Kendrew
  et al.}{2015}]{Kendrew2015}
Kendrew S.  et al., 2015, \mn@doi [PASP] {10.1086/682255}, 127, 623

\bibitem[\protect\citeauthoryear{Kreidberg et al.,}{Kreidberg
  et al.}{2014}]{Kreidberg2014}
Kreidberg L. et al., 2014, \mn@doi[Nature]{10.1038/nature12888}, 505, 69 



\bibitem[\protect\citeauthoryear{Leisenring et al.}{Leisenring et al.}{2018}]{Leisenring2018}
Leisenring J. et al., 2018, SPIE Astronomical Telescopes +
Instrumentation, Austin, Texas, USA, 10-15 June 2018, 10698-184


\bibitem[\protect\citeauthoryear{Leinert et al.}{Leinert et al.}{1998}]{Leinert1998}
Leinert et al.,  1998, \mn@doi [Astron. Astrophys. Suppl. Ser.]{10.1051/aas:1998105}, 127, 1

\bibitem[\protect\citeauthoryear{Line et al.}{Line et al.}{2013}]{Line2013}
Line M. R.  et al., 2013, \mn@doi [ApJ]
  {10.1088/0004-637X/775/2/137}, 775, 137


\bibitem[\protect\citeauthoryear{Madhusudhan \& Seager}{Madhusudhan \&
  Seager}{2009}]{Madhusudhan2009}
Madhusudhan N.,  Seager S.,  2009, \mn@doi [ApJ]
  {10.1088/0004-637x/707/1/24}, 707, 24


\bibitem[\protect\citeauthoryear{Madhusudhan}{Madhusudhan}{2018}]{Madhusudhan2018}
Madhusudhan N.,  2018, In: Deeg H., Belmonte J. (eds), \mn@doi[Handbook of Exoplanets. Springer, Cham]
  {10.1007/978-3-319-55333-7_104}, pp 2153-2182  
 
 
\bibitem[\protect\citeauthoryear{Madhusudhan}{Madhusudhan}{2019}]{Madhusudhan2019}
Madhusudhan N., 2019, {preprint (\mn@eprint {} {arXiv:1904.03190})}

\bibitem[\protect\citeauthoryear{{Mandel} \& {Agol}}{{Mandel} \&
  {Agol}}{2002}]{Mandel2002}
{Mandel} K.,  {Agol} E.,  2002, \mn@doi [ApJ] {10.1086/345520}, 580, L171

\bibitem[\protect\citeauthoryear{McKinney}{McKinney}{2010}]{McKinney2010}
{McKinney} W., 2010, Data Structures for Statistical Computing in Python, \href {http://conference.scipy.org/proceedings/scipy2010/mckinney.html} {Proceedings of the 9th Python in Science Conference}, 51-56  
  
  
\bibitem[\protect\citeauthoryear{Morello}{Morello}{2019}]{Morello2019}
{Morello} G. et al., 2019, {preprint(\mn@eprint {} {arXiv:1908.09599v1}) https://github.com/ucl-exoplanets/ExoTETHyS}

\bibitem[\protect\citeauthoryear{Nielsen et al.}{Nielsen et al.}{2008}]{Nielsen2016}
Nielsen L. D. et al., 2016, \mn@doi[Proc. SPIE]{10.1117/12.2231624}, 99043O 


\bibitem[\protect\citeauthoryear{Oliphant,}{Oliphant}{2007}]{Oliphant2006}
Oliphant T. E., 2006, A guide to NumPy, USA: Trelgol Publishing 
 
  

\bibitem[\protect\citeauthoryear{Parviainen}{Parviainen}{2015}]{Hannu2015}
Parviainen H.,  2015, \mn@doi [MNRAS] {10.1093/mnras/stv894}, 450, 3233

\bibitem[\protect\citeauthoryear{Pascale et al.,}{Pascale
  et al.}{2015}]{Pascale2015}
Pascale E.  et al., 2015, \mn@doi [Exp. Astron.]
  {10.1007/s10686-015-9471-0}, 40, 601

\bibitem[\protect\citeauthoryear{{Perrin et al.}}{{Perrin} et al.}{2012}]{Perrin2012}
Perrin M. D. et al.,  2012, \mn@doi[Proc. SPIE] {10.1117/12.925230}, 84423D

\bibitem[\protect\citeauthoryear{Pirzkal et al.}{Pirzkal
  et al.}{2011}]{Pirzkal2011}
Pirzkal N. et al., 2011, Instrument Science Report WFC3-2011-11, Baltimore, USA: STScI


\bibitem[\protect\citeauthoryear{Piqueras et al.}{Piqueras et al.}{2008}]{Piqueras2008}
Piqueras L. et al.,  2008, \mn@doi[Proc. SPIE]{10.1117/12.789056}, 70170Z


\bibitem[\protect\citeauthoryear{Pont, Zucker  \& Queloz}{Pont
  et al.}{2006}]{Pont2006}
Pont F.,  Zucker S.,   Queloz D.,  2006, \mn@doi [MNRAS] {10.1111/j.1365-2966.2006.11012.x}, 373, 231


\bibitem[\protect\citeauthoryear{Pontoppidan et al.}{Pontoppidan et al.}{2016}]{Pontoppidan2016}
Pontoppidan K. M. et al.,  2016, \mn@doi[Proc. SPIE]{10.1117/12.2231768}, 991016 


\bibitem[\protect\citeauthoryear{Rackham, Apai  \& Giampapa}{Rackham
  et al.}{2018}]{Rackham2018}
Rackham B. V.,  Apai D.,   Giampapa M. S.,  2018, \mn@doi [ApJ] {10.3847/1538-4357/aaa08c}, 853, 122


\bibitem[\protect\citeauthoryear{Rauscher et al.}{Rauscher et al.}{2007}]{Rauscher2007}
Rauscher B. J. et al.,  2007, \mn@doi [PASP] {10.1086/520887}, 119, 768

 
\bibitem[\protect\citeauthoryear{Rein}{Rein}{2012}]{Rein2012}
Rein H.,  2012, {preprint(\mn@eprint {} {arXiv:1211.7121})}

\bibitem[\protect\citeauthoryear{Ressler et al.,}{Ressler
  et al.}{2015}]{Ressler2015}
Ressler M. E.,  et al., 2015, \mn@doi [PASP] {10.1086/682258}, 127, 675

\bibitem[\protect\citeauthoryear{Rieke et al.,}{Rieke
  et al.}{2015a}]{Rieke2015}
Rieke G. H.,  et al., 2015a, \mn@doi [PASP] {10.1086/682252}, 127, 584

\bibitem[\protect\citeauthoryear{Rieke et al.,}{Rieke
  et al.}{2015b}]{Rieke2015b}
Rieke G. H.,  et al., 2015b, \mn@doi [PASP] {10.1086/682257}, 127, 665

\bibitem[\protect\citeauthoryear{Rohatgi}{Rohatgi}{2017}]{Webplot}
Rohatgi A.,  2017, WebPlotDigitizer, \\http://arohatgi.info/WebPlotDigitizer

\bibitem[\protect\citeauthoryear{Sarkar, Papageorgiou  \& Pascale}{Sarkar
  et al.}{2016}]{Sarkar2016}
Sarkar S.,  Papageorgiou A.,   Pascale E.,  2016, 
  \mn@doi[Proc. SPIE]{10.1117/12.2234216}, 99043R 

\bibitem[\protect\citeauthoryear{Sarkar et al.}{Sarkar et al.}{2018}]{Sarkar2018}
Sarkar S. et al., 2018, \mn@doi [MNRAS]
  {10.1093/mnras/sty2453}, 481, 2871

\bibitem[\protect\citeauthoryear{Seager \& Sasselov}{Seager \&
  Sasselov}{2000}]{Seager2000}
Seager S.,  Sasselov D. D.,  2000, \mn@doi [ApJ]
  {10.1086/309088}, 537, 916

\bibitem[\protect\citeauthoryear{{Sing} et al.,}{{Sing}
  et al.}{2016}]{Sing2016}
{Sing} D. K.  et al., 2016, \mn@doi [Nature] {10.1038/nature16068}, 529, 59

\bibitem[\protect\citeauthoryear{{STScI}}{{STScI}}{2017}]{Obs2017}
{Space Telescope Science Institute}, 2016, User
  documentation for Cycle 1: JWST Observatory,  Baltimore, USA: STScI/NASA/ESA/CSA

\bibitem[\protect\citeauthoryear{{STScI}}{{STScI}}{2018a}]{NIRISS2019}
{Space Telescope Science Institute}, 2018a, User
  documentation for Cycle 1: NIRISS,  Baltimore, USA: STScI/NASA/ESA/CSA  

\bibitem[\protect\citeauthoryear{{STScI}}{{STScI}}{2018b}]{NIRCam2019}
{Space Telescope Science Institute}, 2018b, User
  documentation for Cycle 1: NIRCam,  Baltimore, USA: STScI/NASA/ESA/CSA    
  
\bibitem[\protect\citeauthoryear{{STScI}}{{STScI}}{2018c}]{NIRSpec2019}
{Space Telescope Science Institute}, 2018c, User
  documentation for Cycle 1: NIRSpec,  Baltimore, USA: STScI/NASA/ESA/CSA  
  
\bibitem[\protect\citeauthoryear{{STScI}}{{STScI}}{2018d}]{MIRI2019}
{Space Telescope Science Institute}, 2018d, User
  documentation for Cycle 1: MIRI,  Baltimore, USA: STScI/NASA/ESA/CSA


\bibitem[\protect\citeauthoryear{Tinetti et al.,}{Tinetti
  et al.}{2007}]{Tinetti2007}
Tinetti G.  et al., 2007, \mn@doi[Nature]{10.1038/nature06002}, 448, 169 

\bibitem[\protect\citeauthoryear{Tinetti et al.,}{Tinetti
  et al.}{2018}]{Tinetti2018}
Tinetti G.  et al., 2018, \mn@doi[Exp. Astron.]
  {10.1007/s10686-018-9598-x}, 46, 135
  

\bibitem[\protect\citeauthoryear{Tsumura et al.,}{Tsumura
  et al.}{2010}]{Tsumura2010}
Tsumura K. et al., 2010,  \mn@doi [ApJ]
{10.1088/0004-637X/719/1/394}, 719, 394

\bibitem[\protect\citeauthoryear{Varley}{Varley}{2016}]{Varley2016}
Varley R.,  2016, \mn@doi [Comput. Phys. Commun.]
  {https://doi.org/10.1016/j.cpc.2016.05.009}, 207, 298 https://github.com/ryanvarley/ExoData

\bibitem[\protect\citeauthoryear{Waldmann et al.}{Waldmann et al.}{2015}]{Waldmann2015}
Waldmann, I.P. et al.,  2015, \mn@doi[ApJ]
  {10.1088/0004-637x/802/2/107}, 802, 107

  
  

\makeatother
\end{thebibliography}
\bsp	
\label{lastpage}
\end{document}